\documentclass[preprint,showpacs,preprintnumbers,amsmath,amssymb]{revtex4}

\usepackage{graphicx}
\usepackage{dcolumn}
\usepackage{bm}

\begin{document}

\title{Magnetohydrodynamic Stability at a Separatrix: Part II}

\author{Anthony J. Webster}
\affiliation{Euratom/UKAEA Fusion Association, Culham Science Centre,
  Abingdon, Oxfordshire, OX14 3DB.} 
 \email{anthony.webster@ukaea.org.uk}

\date{\today}

 

\noindent

\begin{abstract}
In the first part to this paper\cite{part1} it was shown how a simple
Magnetohydrodynamic model could be used to determine the stability of
a Tokamak plasma's edge to a Peeling (External Kink) mode. 
Stability was found to be determined by the value of $\Delta'$, a
normalised measure of the discontinuity in the radial derivative of
the radial perturbation to the magnetic field at the plasma-vacuum
interface.  
Here we calculate $\Delta'$, but in
a way that avoids the numerical divergences that can arise near a
separatrice's X-point. 
This is accomplished by showing how the method of conformal
transformations may be generalised to allow their application to
systems with a non-zero boundary condition, and using the technique to
obtain analytic expressions for both the vacuum energy and $\Delta'$. 
A conformal transformation is used again to obtain an 
equilibrium vacuum field surrounding a plasma with a separatrix. 
This allows the subsequent evaluation of the vacuum energy and
$\Delta'$.
For a plasma-vacuum boundary that approximates a separatrix, 
the growth rate $\gamma$ normalised by the Aflven frequency $\gamma_A$ 
is then found to have 
$\ln(\gamma/\gamma_A)=-\frac{1}{2} \ln \left( q'/q \right)$. 
Consequences for Peeling mode stability are discussed.

\end{abstract}


\maketitle


\section{Introduction}

Modern Tokamaks and future designs for power-plant scale Tokamaks,
have plasmas with a separatrix at the plasma's edge. 
To determine whether these Tokamak plasmas are susceptible to an ideal
Magnetohydrodynamic Peeling mode (External kink) instability, a simple
model was generalised from a cylindrical to toroidal Tokamak geometry
in the first part\cite{part1}  to this paper. 
The conclusions from part (I) were: 
(a.) Peeling-mode stability is determined by the value of $\Delta'$, a
poloidally averaged measure of the discontinuity in the radial
derivative of the perturbation to the magnetic field at the
plasma-vacuum  interface (the separatrix), and (b.) that regardless of
the sign of $\delta W$ the growth rate could still be vanishingly
small.   
To determine the consequences of part (I) for the stability of the
Peeling mode, this paper evaluates $\Delta'$.   
It is essential that in this calculation any divergences due to the
X-point are incorporated and not accidentally removed or accentuated
by the discretisation of space that is usually required by numerical
modelling, and ideally that $\Delta'$ is calculated exactly.  
That is the purpose of this paper. 
This paper and part (I) are summarised in Ref. \cite{PRL}. 

By using analytical methods to study Peeling mode stability in a
plasma equilibrium with a separatrix boundary (and an X-point), we
hope to avoid difficulties that are encountered with numerical
studies, and to gain a better understanding of the physical factors that 
affect the plasma's stability. 
The techniques developed here may also find further applications to
related problems using different models of the plasma. 
The results from this and future work are intended to provide 
understanding and tests, that will assist the development of codes for
stability calculations that incorporate more advanced models of 
plasma physics and Tokamak geometry.  
It is also hoped that the methods developed in this paper 
will have applications outside of plasma physics.

{\bf Outline}: We review conformal transformations in Section
\ref{CT1}, and describe the Karman-Trefftz transformation\cite{Milne}
in  Section \ref{KT1}.
The Karman-Trefftz transformation provides an example of a
transformation from a circular boundary to a separatrix boundary with
an X-point.   
Section \ref{CP1} reviews how a complex potential may be defined and
used to calculate how the vacuum magnetic field will
transform under a mapping from a system with a circular boundary, to
one with a separatrix.   
For a large aspect ratio system, the vacuum energy is calculated for
both a circular cross-section and a separatrix cross-section in
Section \ref{D}, obtaining the vacuum energy for a separatrix
cross-section in terms of a sum of Fourier co-efficients.  
The Fourier co-efficients are determined by the plasma-vacuum boundary
conditions, and Section \ref{BCsec} discusses how these boundary
conditions are modified by a conformal transformation.  
This is where we have departed from the conventional textbook
applications of conformal transformations, that require the boundary
conditions to be zero.  
Instead the transformed boundary conditions presented in Section
\ref{BCsec} provide analytic expressions to determine the Fourier
co-efficients in terms of the straight field line angle, that is not
yet known.  
The straight field line angle is calculated in terms of an equilibrium
vacuum field and the conformal transformation, in Section \ref{strt1}.
Section \ref{EQ1} calculates an equilibrium vacuum field for both the
circular boundary and the separatrix boundary systems, subsequently
allowing analytic expressions for the safety factor $q$ and the
straight field line angle to be obtained at the plasma-vacuum
boundary.  
At this point all the analytic expressions needed to calculate the
vacuum energy have been obtained.  
Section \ref{More} investigates how other quantities (in addition to
the field and the boundary conditions), change under a
conformal transformation.  
These expressions are tested in Section \ref{recalc}, by
re-calculating the vacuum energy from a surface integral
representation given in the first part to this paper.  
It is noted that this calculation infers the value for $\Delta'$
in terms of the vacuum energy $\delta W_V$.  
Reassured with the results from Section \ref{recalc}, Section
\ref{secDelta'} calculates $\Delta'$ directly, and confirms the same
answer in terms of $\delta W_V$ found in the previous section.  
At this stage we have analytic expressions for $\delta W_V$ and
$\Delta'$ in terms of a sum of Fourier coefficients, and analytic
expressions for the Fourier coefficients. 
Section \ref{Sum1} calculates $\delta W_V$ by evaluating the sum of
Fourier coefficients in a number of ways, finding the same result 
as for an equivalent calculation in a circular cross-section system,
and independent of the details of the separatrix geometry.  
Section \ref{Conc1} provides a discussion that compares and extends
previous work, considers the mode structure, and summarises the paper.

\section{Conformal Transformations}\label{CT1}

A conformal transformation  $w(z)$ (e.g. see Ref. \cite{Spiegel}),
is an analytic function between 
complex planes $z \rightarrow w(z)$. It has the property that the
angle (and direction of the angle), between curves in the plane from
which they are mapped, is retained between the resulting curves in the
plane onto which it is mapped. 
This angle-preserving property ensures that the unit normal to a line
in one plane will map to a vector normal to the mapped line, with a
consequence that a boundary condition of $0=\vec{n}_z.\vec{B}_z$ 
will be transformed to a boundary condition of $0= \vec{n}_w
. \vec{B}_w$ for the boundary in the $w(z)$ plane.  
More generally, if the arc length around a boundary contour in the
$z$-plane is parameterised by $\alpha$, then it will be shown in
Section \ref{BCsec} that  
$\left. \vec{n}_w . \nabla_w V(w) \right|_\alpha = \left|
\frac{dz}{dw} \right| \left. \vec{n}_z . \nabla_z
V(z)\right|_{\alpha}$.  

An important property of conformal maps is that a 
function that satisfies Laplace's equation, 
will continue to do so after a conformal transformation. In
other words, if $\nabla^2 V(z)=0$, then provided $w(z)$ is a conformal
transformation, $\nabla^2_w V(z(w))=0$ also. 
Riemann's mapping theorem indicates the existence of a conformal
transformation from a circle to any closed region. So provided a
suitable transformation may be found, and provided the boundary
conditions map in a simple enough way (as they often do), then it is
possible to
find solutions in complicated geometries by solving a 
problem with a simple circular boundary.

\section{Karman-Trefftz transformation}\label{KT1}

A mapping that may be used to take us from a circle to a shaped cross
 section with an X-point,  
 is the Karman-Trefftz transformation\cite{Milne}. 
The Karman-Trefftz transformation is a generalisation of the Joukowski
transformation that is well known for its use in aerodynamics
calculations for the lift from an airplane wing. It maps from a domain
that surrounds a circle 
containing the point $z=-l$ and whose edge passes through $z=l$, to
w(z), with 
\begin{equation}\label{KTtransform}
\left( \frac{w+n l}{w-nl} \right) = \left( \frac{z+l}{z-l} \right)^n 
\end{equation}
For simplicity we will restrict ourselves to domains of $z$ that are
symmetric about the real axis, so that $w(z)$ is also symmetric about
the real axis with a boundary that has 
\begin{equation}
z=-a +(a+l)e^{i\alpha}
\end{equation}
with $a>0$, so that $\alpha$ will parameterise the boundary curve (in
both the $z$ and $w(z)$  planes).  
Note that $\alpha$ is not the argument of either
$z$ or $w(z)$  (if we are centred on $z=a$ in the z-plane, $\alpha$ is 
the poloidal angle).   
If $n=3/2$, then the cusp-like point at $z=l$ becomes an X-point (with
a $\pi/2$ interior angle at the joining surfaces in the $w(z)$
plane)\cite{Milne}, and $n=2$ produces the Joukowski
transformation.
By making $l/a$ arbitrarily small we make the X-point region
arbitrarily localised, a situation similar to that described by
 Webster\cite{Webster08}.  
This may be seen by rearranging Eq. \ref{KTtransform} to give
\begin{equation}
w(z)=-nl \left( \frac{(z-l)^n + (z+l)^n}{(z-l)^n - (z+l)^n} \right) 
\end{equation}
and  writing it as an asymptotic expansion in $l/z$, with
\begin{equation}
w(z)=z + l \left( \frac{l}{z} \right)\frac{5}{12} \left\{ 1 +
\frac{7}{60} \left( \frac{l}{z} \right)^2 + \frac{13}{300} \left(
\frac{l}{z} \right)^4 + .. \right\} 
\end{equation}
So provided $\left| l/z \right|$ is sufficiently small then $w(z)=z$. 

In summary, the Karman-Trefftz transformation provides an explicit
representation of a transformation from a circular boundary (with
$z=-a+(a+l)e^{i\alpha}$), to a shaped boundary with an X-point. 

\section{The complex potential}\label{CP1}

We will need to know how the vacuum magnetic field (the gradient of a
potential) is transformed as we move from a circular cross-section to
the X-point geometry. 
This is most easily accomplished by 
representing the magnetic field as a complex number whose real and
imaginary parts are interpreted as its $x$ and $y$ components, and by
defining a complex potential $\Omega$ in terms of the magnetic
potential $V$.

The complex representation for the magnetic field is given in terms of
the magnetic potential $V$, with 
\begin{equation}
B_z = \frac{\partial V}{\partial x} + i \frac{\partial V}{\partial y} 
\end{equation}
The complex potential $\Omega$ is defined in terms of $V$ and the
conjugate function of $V$, such that $\Omega$ is analytic and 
satisfies the Cauchy-Riemann  equations. 
Specifically, 
\begin{equation}
\Omega(z)=V(z)+i\psi(z)
\end{equation}
with $\psi$ the conjugate function of $V$. Then the Cauchy-Riemann
conditions are satisfied, with 
\begin{equation}
\begin{array}{l}
\frac{\partial V}{\partial x} = \frac{\partial \psi}{\partial y} 
\\ 
\frac{\partial V}{\partial y} = - \frac{\partial \psi}{\partial x} 
\end{array}
\end{equation}
The Cauchy-Riemann conditions may be used to show that 
\begin{equation}
B_z = \overline{ \frac{d\Omega}{dz} }
\end{equation}
and hence the field in the transformed system $w(z)$ may now easily be
found from 
\begin{equation}\label{Bw} 
B_w = \overline{ \frac{d\Omega}{dw} }
= \overline{ \frac{d\Omega}{dz} \frac{dz}{dw} }
= B_z \overline{ \frac{dz}{dw} }
\end{equation}

\section{The vacuum energy}\label{D}

Working in terms of the complex magnetic field and the
complex potential, we have 
\begin{equation}\label{dWBw}
\delta W_V = \int \left| B_w \right|^2 dw_x dw_y 
\end{equation}
where the integral extends from the boundary that is parameterised by
$\alpha$ at $w(z(\alpha))$, to infinity. 
To evaluate the integral we use Eq. \ref{Bw} so that 
$|B_w|^2 = 
\left| { \frac{d\Omega}{dz} } \right|^2 
\left| { \frac{dz}{dw} } \right|^2 $ 
and we change coordinates back to the circular cross-section
coordinates, with $dw_x dw_y = \frac{\partial (w_x,w_y)}{\partial
  (x,y)}dxdy$ where $ \frac{\partial (w_x,w_y)}{\partial (x,y)} = 
\left|\frac{dw}{dz}\right|^2 $ because $w(z)$ is an analytic
function\cite{Spiegel}.  
So when we change into the $z$ coordinates (for the purpose of
evaluating the integral Eq. \ref{dWBw}), the factors of 
$\left| \frac{dw}{dz}  \right|^2$ and $\left| \frac{dz}{dw}  \right|^2$
cancel to give 
\begin{equation}
\delta W_V = \int \left| B_z \right|^2 dx dy 
\end{equation}
In a similar way it may be shown that $\oint_{l_w} B_w . dl_w =
\oint_{l_z} B_z . dl_z$, reflecting the fact that the same total
current is contained within $l_z$ and $l_w$.

Hence the vacuum energy may be found in terms of the vacuum energy for
a solution with a circular boundary, which is much easier to
calculate. The actual values of $\delta W_V$ and $B_z$ are determined
by the plasma-vacuum boundary conditions, that will be modified by the
transformation. The mapping of the boundary condition and the
resulting boundary condition are obtained in Section \ref{BCsec}, but
for the present we will obtain the general solution in terms of the
Fourier coefficients that the boundary conditions will determine.

The vacuum field has $\nabla \wedge \vec{B}_V = \nabla . \vec{B}_V
=0$, so we may write $\vec{B}_V=\nabla V$ with $\nabla^2 V=0$.
We will start from a co-ordinate system with a circular cross section toroidal
geometry, then subsequently obtain a 2D problem by taking the large
aspect ratio limit. In this co-ordinate system  
\begin{equation}
\nabla^2 V = \frac{1}{r} \frac{\partial}{\partial r} r\frac{\partial
  V}{\partial r} + \frac{1}{r^2} \frac{\partial^2 V}{\partial
  \alpha^2} + \frac{1}{R^2} \frac{\partial^2 V}{\partial \phi^2}  
\end{equation}
with $r$, $\alpha$, $\phi$ the radial coordinate, poloidal and
toroidal angle respectively. 
Writing 
\begin{equation}\label{Vexp}
V = \sum_{p=-\infty}^{p=\infty} e^{ip\alpha - in\phi} V_p(r) 
\end{equation}
and then projecting out the Fourier components, requires  
\begin{equation}
0 = \frac{1}{r} \frac{\partial }{\partial r} r \frac{\partial
  V_p}{\partial r} - p^2 V_p(r) - n^2 \left( \frac{r}{R} \right)^2
  V_p(r) 
\end{equation}
which for any given (finite) $n$ becomes 2-dimensional when $r/R\rightarrow 0$,
leaving  
\begin{equation}
0 = r^2 \frac{\partial^2 V_p}{\partial r^2} + r \frac{\partial
  V_p}{\partial r} - p^2 V_p 
\end{equation}
that has solutions that tend to zero as $r\rightarrow \infty$, of $V_p
= a_p \left(\frac{r_a}{r} \right)^{|p|}$, where the $a_p$ will be
determined by the boundary conditions, and $r_a$ denotes the radial
position of the plasma-vacuum surface.   
Note that it is only because the large aspect ratio 
limit makes the problem 2-dimensional, that we are able to use
conformal transformations in the calculation.

The vacuum energy is 
\begin{equation}\label{dWV1}
\delta W_V 
=\frac {1}{2} \int \left| B_1^V \right|^2 \vec{dr}
\end{equation}
Into which we now substitute $V$, with 
\begin{equation}\label{Vvac}
V = \sum_{p=-\infty}^{p=\infty} e^{ip\alpha - in\phi} a_p
\left(\frac{r_a}{r}\right)^{|p|}  
\end{equation}
and integrate with respect to $\phi$ and $\alpha$, with  
$r=r_a$ at the surface, to get 
\begin{equation}\label{FdWV}
\delta W_V = 2\pi^2 R \sum_{p\neq0}  |p| \left| a_p \right|^2 
\end{equation}
Here and in the remainder of this article $R$ will be taken as a
typical measure of the major radius that is approximately constant and
independent of the poloidal angle.  
$R$ is identical to the $R_0$ of the first part to this paper, but
because the rest of this paper considers a large aspect ratio limit,
we will simply write $R$ as opposed to $R_0$.

\section{Boundary conditions}\label{BCsec}

The boundary conditions in the $w$-plane determine $\vec{n}.\vec{B}$
in terms of the plasma perturbation. 
We will write $\vec{n}.\vec{B}$ in the $w$-plane as $n_w.B_w$.
However, to obtain the co-efficients $a_p$ we need to know
$\vec{n}.\vec{B}$ in  
the $z$-plane (that we write as $n_z.B_z$). 
Therefore we need to know how $\vec{n}.\vec{B}$ is transformed as we
map between  
the  $z$-plane where the boundary is circular with $z(\alpha)=-a
+(a+l)e^{i\alpha}$,  
and the $w$-plane whose boundary is shaped and contains an X-point. 
This is calculated next.

Recall that the real and imaginary components are considered as
orthogonal vector components. Then the tangent vector $t_w(\alpha)$ of
the surface traced by $w(z(\alpha))$ is simply given by  
\begin{equation}
t_w(\alpha) = \frac{ 
\frac{\partial w(z(\alpha )) }{\partial \alpha} }{ 
\left| \frac{\partial w}{\partial \alpha} \right|} 
\end{equation}
However, using the fact that $w(z)$ is an analytic function, so that
$\frac{\partial w}{\partial \alpha} =\frac{dw}{dz} \frac{\partial
  z}{\partial \alpha}$, then 
\begin{equation}
t_w(\alpha) = \frac{\frac{\partial w(z(\alpha))}{\partial
    \alpha}}{\left|\frac{\partial w}{\partial \alpha }\right|}  
=
    \frac{\frac{dw}{dz(\alpha)}
\frac{\partial z(\alpha)}{\partial \alpha}}
{\left|\frac{dw}{dz(\alpha)}\right|\left|
    \frac{\partial z}{\partial \alpha} \right|}  
= \frac{\frac{dw}{dz}}{\left| \frac{dw}{dz} \right|}t_z 
\end{equation}
where the tangent $t_z$ of $z(\alpha)$ in the $z$-plane is again
simply given by $t_z(\alpha)=\frac{\partial z}{\partial \alpha}/\left|
\frac{\partial z}{\partial \alpha} \right|$.  
To obtain the unit normals we rotate the tangent vector by $-\pi/2$,
by simply multiplying by $-i$. Hence  
\begin{equation}
n_w=-it_w=\left(\frac{w'(z)}{|w'(z)|}\right)(-it_z)=\frac{w'(z)}{|w'(z)|}n_z
\end{equation}

We already know that $B_w = B_z \overline{\frac{dz}{dw}}$, so to obtain how
$n_w.B_w$ transforms we need to simplify  
\begin{equation}
n_w.B_w = \left( \frac{w'(z)}{|w'(z)|} n_z \right) . \left( B_z
\overline{\frac{dz}{dw}} \right)  
\end{equation}
where the dot product refers to the sum of the product of the real
parts, plus the product of the imaginary parts (examples may be found
in Appendix \ref{complexidentities}). 
We will use $\overline{\frac{dz}{dw}}= \overline{1/\frac{dw}{dz}} 
= \frac{dw}{dz}/\left| \frac{dw}{dz}\right|^2$, and write
$n_z=e^{i\theta_z}$, $B_z=r_Be^{i\theta_B}$, and
$\frac{dw}{dz}=r_{w'}e^{i\theta_{w'}}$, to give  
\begin{equation}
\begin{array}{ll}
n_w.B_w 
&= \frac{w'(z)}{|w'(z)|} n_z . B_z \frac{w'(z)}{|w'(z)|^2} 
\\
&= \frac{r_B}{r_{w'}} \left[ e^{i\theta_{w'}} e^{i\theta_z}
  . e^{i\theta_B}e^{i\theta_{w'}}  \right]  
\\
&= \frac{r_B}{r_{w'}} \left[ \cos(\theta_{w'} + \theta_z) + i \sin(
  \theta_{w'} + \theta_z ) \right]  . 
\left[ \cos(\theta_{w'}+\theta_B) + i \sin(\theta_{w'} + \theta_B)     
 \right]
\\
&= \frac{r_B}{r_{w'}} \left[ \cos(\theta_{w'} + \theta_z)
  \cos(\theta_{w'} + \theta_B)  
\right. + \left. 
\sin( \theta_{w'} + \theta_z ) \sin( \theta_{w'} + \theta_B ) \right]
\\
&= \frac{r_B}{r_{w'}} \cos( \theta_z - \theta_B ) 
\\
&= \frac{r_B}{r_{w'}} e^{i\theta_z} . e^{i\theta_B} 
\\
&= \frac{1}{r_{w'}} n_z . B_z 
\\
&= \frac{n_z.B_z}{\left| \frac{dw}{dz} \right|}  
\end{array}
\end{equation}
This calculation is repeated by an alternative method in Section \ref{More}.

Knowing how $\vec{n}.\vec{B}$ transforms between $z$ and the $w(z)$
plane, we now return to the plasma-vacuum boundary conditions.  
As shown in part (I), the plasma-vacuum boundary condition is 
\begin{equation}\label{EPSBC}
\left. \nabla \psi . \vec{B} \right|_{edge} = 
\left. \nabla \psi . \vec{B}^V \right|_{edge}
\end{equation}
where "edge", refers to the equilibrium position of the surface. 
Because $\vec{B}_0 . \nabla \xi_{\psi}=\nabla \psi . \vec{B}_1$, we
therefore require that  
\begin{equation}
\left. \vec{B}_0. \nabla \xi_{\psi} \right|_{edge} = 
\left. \nabla   \psi . \vec{B}_1^V  \right|_{edge}
\end{equation}
For a single Fourier mode in straight field line co-ordinates
$\xi_{\psi}=\xi_m(\psi)e^{im\theta}$, 
with $\theta=\frac{1}{q}\int^{\chi} \nu d\chi'$, $q=\frac{1}{q} \oint
\nu d\chi'$, $\nu =\frac{IJ_{\chi}}{R^2}$, $J_{\chi}$ the Jacobian of
the orthogonal $\chi$, $\psi$, $\phi$ co-ordinate system, with $\psi$
the poloidal flux, $\chi$ the poloidal angle, and $\phi$ the toroidal
angle, and $I(\psi)$ is the flux function for which $\vec{B}= I \nabla
\phi + \nabla \phi \wedge \nabla \psi$. 
After taking derivatives $\vec{B} . \nabla \xi_{\psi}$ then gives 
\begin{equation}
\nabla \psi . \vec{B}_1^V = \left( im - inq \right) \frac{I}{qR^2}
\xi_m e^{im\theta -in\phi}  
\end{equation}
This may alternately be written as 
\begin{equation}
n_w . B_w = im\Delta \frac{\xi_m}{RB_p} \frac{I}{qR^2} e^{im\theta -in\phi} 
\end{equation}
where $\Delta=\frac{m-nq}{m}$.


Now we transform into the $z$ coordinates, transforming both
$n_w.B_w=n_z.B_z/|w'(z)|$ and $B_p=B_{pz}/|w'(z)|$, to get 
\begin{equation}\label{nzBC}
n_z . B_z = im\Delta \frac{\xi_m}{R} \frac{|w'(z)|^2}{|B_{pz}|}
\frac{I}{qR^2} e^{im\theta -in\phi}  
\end{equation}
Using Eq. \ref{Vexp} along with $V_k = a_k \left( \frac{r_a}{r}
\right)^{|k|}$, and that $n_z.B_z = \vec{e}_r . \nabla V$, then gives  
\begin{equation}
\sum_{k=-\infty}^{\infty} a_k \frac{-|k|}{r_a} e^{ik\alpha}  =
im\Delta \frac{\xi_m}{R} \frac{|w'(z)|^2}{|B_{pz}|} \frac{I}{qR^2}
e^{im\theta}  
\end{equation}
From which the Fourier coefficients are easily obtained by multiplying
by $\frac{e^{-ip\alpha}}{2\pi}$ and integrating from $\alpha = -\pi$
to $\pi$, to give   
\begin{equation}\label{ap1}
a_p = - \left(\frac{\Delta}{|p|} \right) \frac{\xi_m}{R}
\frac{1}{2\pi} \oint im  \frac{|w'(z)|^2}{|B_{pz}|} \frac{Ir_a}{qR^2}
e^{im\theta -ip\alpha} d\alpha  
\end{equation} 

In the following section we will see that 
\begin{equation}
\theta(\alpha)=\frac{Ir_a}{qR^2} \int^{\alpha}
\frac{|w'(z)|^2}{|B_{pz}|} d\alpha 
\end{equation} 
which will allow us to integrate by parts once to get 
\begin{equation}\label{ap}
a_p = - \left(\frac{ip}{|p|} \right) \Delta \frac{\xi_m}{R}
\frac{1}{2\pi} \oint e^{im\theta -ip\alpha} d\alpha  
\end{equation}
To evaluate the coefficients $a_p$, we will need an expression for
$\theta(\alpha)$, this is addressed in the following 2 Sections.

\section{The straight field-line angle}\label{strt1}

In the absence of equilibrium skin currents, 
the plasma's equilibrium field $\vec{B}_0$ equals the vacuum's
equilibrium field $\vec{B}_0^V$ at the surface between the plasma and
the vacuum, with  
$\left. \vec{B}_0 \right|_{edge}= \left. \vec{B}_0^V \right|_{edge}$.  
Therefore provided that we know the equilibrium 
vacuum field at the surface, then we also know the plasma's
field at the surface. Consequently, if we know the vacuum field at the
surface, then it is possible to calculate the straight field-line
variable at the surface. Firstly we note that  
\begin{equation}\label{sfl1}
\theta = \frac{1}{q}\int^{\chi} \nu d\chi
=\frac{I}{qR^2} \int^{\chi} \frac{J_{\chi}B_p d\chi}{B_p}
=\frac{I}{qR^2} \int^{l} \frac{dl}{B_p}
\end{equation}
An element of arc length parallel to the tangent vector, $dl_w$ has 
\begin{equation}
dl_w = \frac{\partial w}{\partial \alpha} d\alpha
= \frac{dw}{dz} \frac{\partial z}{\partial
  \alpha} d\alpha  
= \frac{dw}{dz} dl_z
\end{equation}
Hence an element of arc length $|dl_w|$ transforms such that
$|dl_w|=\left|\frac{dw}{dz}\right|  |dl_z|= \left| \frac{dw}{dz}
\right| r_a d\alpha$, for a circular cross section of radius $r_a$ in
the $z$-plane.  
Using this plus $|B_w|=|B_z|/|w'(z)|$, we may write Eq. \ref{sfl1}  as 
\begin{equation}\label{theta1}
\theta(\alpha) =
\frac{I}{qR^2} \int^{l_w} \frac{|dl_w|}{|B_{pw}|}
=\frac{Ir_a}{qR^2} \int^{\alpha} \left| \frac{dw}{dz} 
\right|^2 \frac{d\alpha}{|B_{pz}|} 
\end{equation}
Hence if we know the equilibrium field, then we can obtain an analytical
expression for the straight field-line coordinate as a function of
$\alpha$ in the $z$-plane.

\section{Equilibrium vacuum field}\label{EQ1}

\begin{figure}[tpb!]
\begin{center}
\resizebox{125mm}{!} 
{\includegraphics[angle=0.]{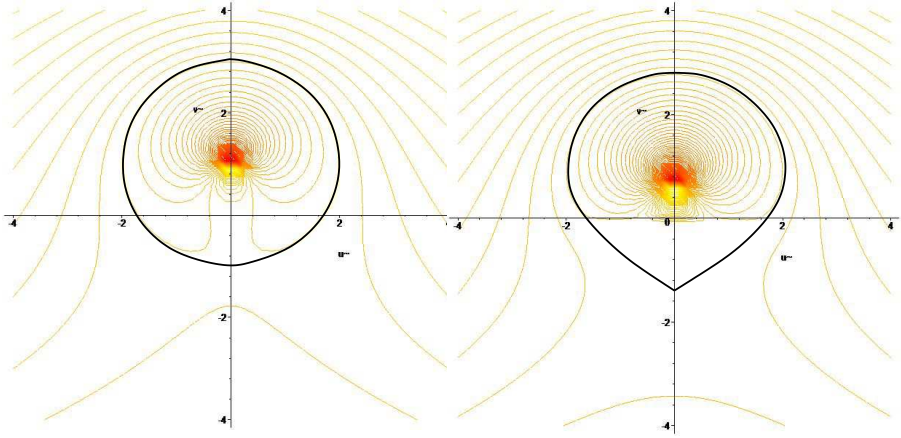}}
\caption{
The figure shows contour plots of the imaginary part of the complex
potential Eq. \ref{omegaz} for the magnetic field, with $a=l=1$.  
Such plots give streamlines of the magnetic field\cite{Spiegel}. 
The plot on the left is equivalent to a
combination of a vertical field and that produced by a current through 
$z=i$, with a 
field strength such that $B_z=0$ at the bottom of the plasma vacuum
surface.
The plot on the right is obtained from $\Omega(z)$ by a conformal
transformation, writing $\Omega(z(w))$ and calculating plots of
$\Omega(z(w))$ in the $w$-plane.  
$z(w)$ is obtained from Eq. \ref{KTtransform} with $n=3/2$ and $l=1$,
that gives  
$z(w) = \frac{(w+3/2)^{2/3} + (w-3/2)^{2/3}}{(w+3/2)^{2/3} - (w-3/2)^{2/3}}$. 
For the plots the domain of $z$ and $w$ were rotated by substituting
with $iz$ and $iw$ respectively, so that the X-point is seen at the 
bottom of the figure. 
}\label{magneticfieldplots}  
\end{center}
\end{figure}

The equilibrium vacuum field must (i) have a potential that satisfies
Laplace's equation in the vacuum region, (ii) have $\vec{n}
. \vec{B}_0=0$ at the plasma-vacuum boundary (including at the 
strongly shaped X-point containing equilibrium), and (iii) have the
field $B_0=0$ at the X-point. The first part is most easily satisfied
- we can take a solution that satisfies Laplace's equation and
$n_z.B_z=0$ for a circular cross section, and after a conformal
transformation to a shaped cross-section we will still have
$n_w.B_w=0$ and a potential that satisfies Laplace's equation.  
To obtain a field with $B_p=0$ at the X-point, we follow a procedure
that is equivalent to that when applying the Kutta condition to obtain
the flow around an airplane wing (using a conformal transformation).  
Essentially, in the z-plane we combine  a homogeneous
horizontal field and a circulating field, such that the field becomes
zero at a single point on the circular boundary. This is physically
equivalent to imposing an external horizontal field, and then driving
a current through the plasma. Mathematically it corresponds to taking
a complex potential in the $z$-plane of  
\begin{equation}\label{omegaz}
\Omega = B_{p0}\left\{ i(z+a) - i \frac{(a+l)^2}{(z+a)} -2i(a+l) \ln
(z+a) \right\}  
\end{equation} 
with a boundary at 
\begin{equation}\label{zsep}
z=-a +(a+l)e^{i\alpha} 
\end{equation}
with $\alpha \in [0,2\pi]$, $B_{p0}$ a dimensional constant, and the
radius of the circular boundary $r_a = (a+l)$.  
The sign of $B_{p0}$ determines the direction of the circulation of
$B_{pz}$, clockwise ($B_{p0}<0$) or anticlockwise ($B_{p0}>0$), and
the field $B_{pz}$ is obtained from  
$B_{pz}=\overline{\frac{d\Omega}{dz}}$, with 
\begin{equation}
B_{pz} = \overline{\frac{d\Omega}{dz}} = i\frac{(z-l)^2}{(z+a)^2} B_{p0} 
\end{equation}
which at the separatrix given by Eq. \ref{zsep} gives $\left| B_{pz}
\right|= 2B_{p0} (1-\cos(\alpha) )$.  
To consider an outermost flux surface that is just inside the
separatrix, we may instead consider  
\begin{equation}
z = -a + (a+l-\epsilon)e^{i\alpha} 
\end{equation}
with $\epsilon \ll l$. Then for $\epsilon \ll l \ll a$ we have 
\begin{equation}\label{Bpzalpha}
\left| B_{pz} \right| \simeq 2 B_{p0}  \left( 1 - \cos(\alpha) +
\frac{\epsilon^2}{2a^2} \right)  
\end{equation}
so instead of $B_{pz}=0$ at the X-point (that is located at
$\epsilon=0$ and $\alpha=0$), we have $B_{pz}=\epsilon^2/a^2$.  
Notice that we have retained the singular perturbation in $\epsilon$
(singular in that although formally $\epsilon^2/a^2 \ll \epsilon/a$, 
it is the term in $\epsilon^2/a^2$ that qualitatively alters
$|B_{pz}|$ by preventing it from being zero), further details are
given in Appendix \ref{w'xpt}.  
The field in the transformed space is given by $B_{pw}= B_{pz}
\overline{\frac{dz}{dw}}$, although we shall not need this here. Plots of
the equilibrium are given in Figure \ref{magneticfieldplots}.  


As we approach the separatrix the behaviour of $\theta(\alpha)$ is
dominated by the  
zeros in $w'(z)$ and $B_{pz}$ that occur near the X-point. Near the
X-point it may be shown (in appendix \ref{w'xpt}), that for the case
of $n=3/2$ with field lines crossing perpendicularly to each other,
that $|w'(z)|^2$ is given by  
\begin{equation}\label{w'zsq}
\left| w'(z) \right|^2 \simeq \frac{1}{\sqrt{2}}\left(\frac{3}{2}
\right)^4 \frac{a}{l}  
\sqrt{ 1 - \cos(\alpha) + \frac{\epsilon^2}{2 a^2} } 
\end{equation}
We may use Eqs. \ref{Bpzalpha} and \ref{w'zsq} to calculate $q$, with  
\begin{equation}
q=\frac{1}{2\pi}\oint \nu d\chi
=\frac{1}{2\pi} \frac{Ir_a}{R^2} 
\oint 
\frac{|w'(\alpha )|^2}{|B_{pz}|} d\alpha
= \frac{1}{2\pi} \frac{Ir_a}{R^2B_{p0}} 
\frac{1}{\sqrt{c_a}} \oint 
\frac{d\alpha}{ \sqrt{ 1-\cos{\alpha} + \epsilon^2/2a^2}} 
\end{equation}
and $\sqrt{c_a}=2\sqrt{2}\left(\frac{2}{3}\right)^4 \left(\frac{l}{a}
\right)$.  
Similarly for $\theta(\alpha)$ we have from Eqs. \ref{theta1},
\ref{Bpzalpha}, and \ref{w'zsq}, that  
\begin{equation}
\theta(\alpha) = \frac{1}{q} \frac{Ir_a}{R^2B_{p0}} 
\frac{1}{\sqrt{c_a}} \int^{\alpha}_{-\pi}  
\frac{d\alpha}{ \sqrt{ 1-\cos{\alpha} + \epsilon^2/2a^2}} 
\end{equation} 
Because the integral is dominated by the divergence at the X-point
where $\alpha=0$, $q$ may be approximated by  
\begin{equation}\label{thetaapprox}
q \simeq \frac{1}{2\pi} \frac{Ir_a}{R^2B_{p0}} \frac{1}{\sqrt{c_a}} \oint 
\frac{d\alpha}{ \sqrt{ \frac{\alpha^2}{2}  + \frac{\epsilon^2}{2a^2} }} 
\simeq \frac{c_t\sqrt{2}}{2\pi} 2 \ln \left( \frac{2a\pi}{\epsilon} \right) 
\end{equation}
with $c_t \equiv \frac{Ir_a}{R^2B_{p0}}\frac{1}{\sqrt{c_a}}$, and similarly 
\begin{equation}
\theta(\alpha) \simeq \frac{c_t}{q} \int^{\alpha}_{-\pi} 
\frac{d\alpha}{ \sqrt{ \frac{\alpha^2}{2}  + \frac{\epsilon^2}{2a^2}
}} 
=\frac{c_t\sqrt{2}}{q} \ln \left(
\frac{ \frac{a\alpha}{\epsilon} + \sqrt{\frac{a^2\alpha^2}{\epsilon^2}
    +1} }{ -\frac{a\pi}{\epsilon} + \sqrt{ \frac{a^2\pi^2}{\epsilon^2}
    + 1} }  
\right)
\end{equation}
Therefore as we approach the separatrix, with $\epsilon \rightarrow
0$, $q$ has a logarithmic divergence with $q \sim -\ln(\epsilon)$ that
is 
typical for a Tokamak plasma near the separatrix. Hence the
qualitative features of $\theta(\alpha)$ 
could have reasonably been postulated without showing that they also
arise from a vacuum field whose potential satisfies Laplace's
equation, has $n_w.B_{pw}=0$ on the separatrix, and with
$B_{pz}\rightarrow 0$ at the X-point; but it is reassuring to know
that this is also the case.

It is interesting to note that in this model for the equilibrium
field, the angle at which the field lines meet at the X-point
determines how strongly the divergence is there.  
For example, if instead of meeting at $\pi/2$ the lines make a cusp
(tending to parallel as they meet), then $q$ is finite
(A cusp is obtained by taking $n=2$ in the Karman-Trefftz
transformation, Eq. \ref{KTtransform}).

\section{More transformed quantities}\label{More}

Now we start to return to our problem of calculating $\Delta'$, by
firstly calculating $\delta W_V$ from  
its surface integral representation that is given in part (I), with  
\begin{equation}\label{EPSdWV}
\delta W_V = \pi \oint J_{\chi} d\chi \left( \frac{i}{n} \right)
\frac{\nabla \psi .\vec{B}_1^*}{B^2} \left[ 
R^2 B_p^2 \frac{i}{n} \frac{\partial}{\partial \psi} \left( \nabla
\psi .\vec{B}_1 \right)  
\right]
\end{equation}
where the integral is over the plasma surface. This 
requires us to know how $\vec{n}.\nabla$ transforms.    
The calculation is done partly to reassure us that the transformed
quantities are correct, but also because it is a simple step to
subsequently obtain $\Delta'$.

First we calculate how $\nabla_w$ transforms. We have 
\begin{equation}
\nabla_z f(z) = \left( \frac{\partial}{\partial x} + i
\frac{\partial}{\partial y} \right) f(z)   
\end{equation}
and 
\begin{equation}
\nabla_w f(z(w)) = \left( \frac{\partial}{\partial w_x} + i
\frac{\partial}{\partial w_y} \right)  f(z(w))   
\end{equation}
where $z=x+iy$ and $w=w_x + iw_y$. 
Now we will use the chain rule to expand $\nabla_z f(z(w))$, noting
that because $w(z)$ is an analytic function it satisfies the
Cauchy-Riemann equations,   
\begin{equation}
\begin{array}{c}
\frac{\partial w_x}{\partial x} = \frac{\partial w_y}{\partial y}
\\
\frac{\partial w_x}{\partial y} = - \frac{\partial w_y}{\partial x}
\end{array}
\end{equation}
and in addition, $\frac{\partial w}{\partial x}=\frac{dw}{dz}$.  
This gives,
\begin{equation}
\begin{array}{ll}
\nabla_z f(z(w)) &= 
\frac{\partial f}{\partial w_x} \frac{\partial w_x}{\partial x} +
\frac{\partial f}{\partial w_y} \frac{\partial w_y}{\partial x} + i
\frac{\partial f}{\partial w_x} \frac{\partial w_x}{\partial y} +i
\frac{\partial f}{\partial w_y} \frac{\partial w_y}{\partial y}  
\\
&= \frac{\partial w_x}{\partial x} \left(
\frac{\partial f}{\partial w_x} + i \frac{\partial f}{\partial w_y} 
\right)
+ \frac{\partial w_y}{\partial x} \left( -i \frac{\partial f}{\partial
  w_x} + \frac{\partial f}{\partial w_y} \right)  
\\
&= \frac{\partial w_x}{\partial x} \left(
\frac{\partial f}{\partial w_x} + i \frac{\partial f}{\partial w_y} 
\right)
- i \frac{\partial w_y}{\partial x} \left( \frac{\partial f}{\partial
  w_x} + i \frac{\partial f}{\partial w_y} \right)  
\\
&= \left( \frac{\partial}{\partial x} \left(w_x - iw_y \right) \right)
\left( \frac{\partial f}{\partial w_x} + i \frac{\partial f}{\partial
  w_y} \right)  
\\
&= \overline{\frac{\partial w}{\partial x}} \nabla_w f
= \overline{\frac{dw}{dz}} \nabla_w f
\end{array}
\end{equation}
Hence we have $\nabla_w = \overline{\frac{dz}{dw}} \nabla_z$. 

Now we consider the transformation of $\vec{n}.\nabla$.
In calculating how more complicated expressions transform, the author
has found the following identities useful, whose derivations are given
in Appendix \ref{complexidentities}.  
\begin{equation}
a.bc=a\bar{b}.c=\bar{a}b.\bar{c}
\end{equation}
\begin{equation}\label{id2}
ab.cd = (a.c)(b.d) + (ia.c)(b.id)
\end{equation}
For example, to calculate $n_w . B_w$ we use Eq. \ref{id2} along with
the results of Section \ref{BCsec} to give  
\begin{equation} 
\begin{array}{ll}
\vec{n}.\vec{B} = n_w.B_w 
&= \left( \frac{w'(z)}{|w'(z)|}n_z \right) . \left(
\overline{\frac{dz}{dw}} 
B_z\right)  
\\
&= \frac{1}{|w'(z)|} \left[
\left(n_z.B_z \right) \left(\frac{dw}{dz}.
\overline{\frac{dz}{dw}} \right)
+ \left(in_z
.B_z\right)\left(\frac{dw}{dz}.i
\overline{\frac{dz}{dw}}\right) 
\right]
\\
&=\frac{n_z.B_z}{\left| w'(z) \right|} 
\end{array}
\end{equation}
as before. In the last step we used,
\begin{equation}
\begin{array}{c}
\frac{dw}{dz}.\overline{\frac{dz}{dw}}=1
\\
\frac{dw}{dz}.i\overline{\frac{dz}{dw}}=0
\end{array}
\end{equation}
that may easily be confirmed by writing $\frac{dw}{dz}=\alpha+i\beta$,
so that $\overline{\frac{dz}{dw}}=\overline{1/(\alpha+i\beta)}$, and
multiplying out.   
A similar calculation for $\vec{n}.\nabla$ gives  
\begin{equation}
\begin{array}{ll} 
\vec{n}.\nabla = n_w.\nabla_w
&= \frac{w'(z)}{|w'(z)|} n_z . \overline{\frac{dz}{dw}} \nabla_z
\\
&= \frac{1}{|w'(z)|} \left[ 
\left( \overline{\frac{dz}{dw}} . \frac{dw}{dz} \right) \left( n_z
. \nabla_z \right)  
+ \left(i\overline{\frac{dz}{dw}} . \frac{dw}{dz} \right) \left( n_z
. i\nabla_z \right)  
\right]
\\
&= \frac{n_z.\nabla_z}{|w'(z)|}
\end{array}
\end{equation}

\section{Recalculating $\delta W_V$}\label{recalc} 

Firstly we re-express Eq. \ref{EPSdWV} using our transformed
quantities, then we show this gives us the same result, Eq. \ref{FdWV}
from Section \ref{D}, before showing how the calculation easily
generalises to give us $\Delta'$ in terms of $\delta W_V$.   

Using 
\begin{equation}
\left. \nabla \psi . \vec{B}_1^V \right|_{edge} = \left. \nabla \psi
. \vec{B}_1 \right|_{edge} = \left. \vec{B} . \nabla \xi_{\psi}
\right|_{edge}  
\end{equation}
gives us 
\begin{equation}
\left. \nabla \psi . \vec{B}_1^V \right|^*_{edge} = - \xi^*_m(\psi_a)
\left( \frac{I}{qR^2} \right) (im) \left( \frac{m-nq}{m} \right)
e^{-im\theta +in\phi}  
\end{equation}
This is substituted into Eq. \ref{EPSdWV}, to give 
\begin{equation}
\delta W_V = \pi \Delta \oint J_{\chi} d\chi \left( \frac{m}{nq}
\right) \left( \frac{I}{R^2B^2} \right) 
\xi_m^* e^{-im\theta +in\phi} \left[ R^2 B_p^2 \frac{i}{n}
  \frac{\partial}{\partial \psi} \nabla \psi . \vec{B}_1 \right]  
\end{equation}
where $\Delta = \frac{m-nq}{m}$. Then using 
$\frac{\partial}{\partial \psi} = \frac{\vec{n}.\nabla}{RB_p} $, 
we get
\begin{equation}
\begin{array}{l}
\delta W_V = \pi \Delta \left( \frac{m}{nq} \right)  \frac{\xi_m^*}{R} 
\oint dl \left( \frac{I}{B^2} \right) 
 e^{-im\theta +in\phi} \frac{i}{n} \vec{n}.\nabla \left( RB_p \vec{n}
 . \vec{B}_1 \right)  
\end{array}
\end{equation}
where we used $dl=J_{\chi}B_p d\chi$. 
Now we will transform this equation into coordinates in which the
plasma has a circular cross-section, using: 
\begin{equation}\label{transforms}
\begin{array}{l}
dl_w = \left| w'(z) \right| dl_z = \left| w'(z) \right| r_a d\alpha 
\\
\left| B_{pw} \right| = \frac{ \left| B_{pz} \right| }{ \left| w'(z) \right| } 
\\
n_w . B_w = \frac{n_z . B_z}{\left| w'(z) \right|} 
\\
n_w . \nabla_w = \frac{n_z . \nabla_z}{\left| w'(z) \right|}  
\end{array}
\end{equation}
After using the chain rule to expand the term in $n_z.\nabla_z$, we
have 
\begin{equation}\label{dWV4}
\begin{array}{l}
\delta W_V = \pi \Delta \left( \frac{m}{nq} \right) \xi_m^* \oint
d\alpha \frac{r_aI}{B^2} e^{-im\theta(\alpha) +in\phi}  
\\
\frac{i}{n} \left[ 
\frac{\left| B_{pz} \right| }{\left| w'(z) \right|^2} n_z . \nabla_z 
\left( n_z . B_z \right) +
n_z . B_z n_z . \nabla_z \left(
\frac{\left|B_{pz}\right|}{\left|w'(z)\right|^2} \right)  
\right]
\end{array}
\end{equation}
The $n_z.\nabla_z$ operator acting on $n_z.B_z$ will produce a term of
order $n$ larger than $n_z.B_z$. Thus usually we would neglect the
second term. However here we need to be careful that there are no
geometrically driven divergences, this is done in Appendix \ref{app1},
where it is confirmed that the term is of order $\frac{1}{n}$ smaller
and may be neglected.  
Hence if we retain only the leading order term in $n$, then
rearranging the expression slightly we have,  
\begin{equation}
\delta W_V = \pi \Delta \left( \frac{m}{nq} \right)^2  \xi_m^* \oint
d\alpha e^{-im\theta(\alpha) +in\phi}  
\frac{r_aI}{B^2} 
\frac{iq}{m} 
\frac{\left| B_{pz} \right| }{\left| w'(z) \right|^2} n_z . \nabla_z
\left( n_z . B_z \right)  
\end{equation}
After comparison with Eq. \ref{theta1} for $\theta(\alpha)$, this may
be written as  
\begin{equation}
\delta W_V = \pi \Delta \left( \frac{m}{nq} \right)^2  \xi_m^* 
\left(\frac{r_a^2I^2}{R^2B^2} \right) 
\oint d\alpha  
\frac{-e^{-im\theta(\alpha) +in\phi}}{im\theta'(\alpha)} 
 n_z . \nabla_z \left( n_z . B_z \right) 
\end{equation}
Using Eq. \ref{Vvac},
\begin{equation}
n_z. \nabla_z \left( n_z . B_z \right) = \left. \frac{\partial^2
  V}{\partial r^2} \right|_{r=r_a}   
= \sum_{p\neq 0} e^{ip\alpha -in\phi} a_p \frac{|p|(|p|+1)}{r_a^2} 
\end{equation}
Giving 
\begin{equation}
\delta W_V = \pi \Delta \left( \frac{m}{nq} \right)^2  \xi_m^* 
\left(\frac{r_a^2I^2}{R^2B^2} \right) 
\sum_{p\neq 0} a_p \frac{|p|(|p|+1)}{r_a^2} 
\oint d\alpha \mbox{ } 
\frac{-e^{-im\theta(\alpha)+ip\alpha} }{im\theta'(\alpha)} 
\end{equation}

Next we observe that 
\begin{equation}\label{keyeq}
\oint d\alpha \frac{1}{im\theta'(\alpha)} e^{ip\alpha -im\theta(\alpha)}
= \oint d\alpha \frac{1}{ip} e^{ip\alpha -im\theta(\alpha)}+ \mbox{O} \left( 
\frac{\left(\frac{\epsilon}{a}\right) \ln \left( \frac{\epsilon}{a}
  \right) }{n} \right) 
\end{equation}
that we will justify below, and that appears to be the key result
linking the high $n$ and $q$ calculations at arbitrary cross-section,
to the circular cross section result.  
Integrating by parts we get
\begin{equation}
\oint d\alpha \frac{e^{ip\alpha -im\theta(\alpha)}}{im\theta'(\alpha)} 
= \oint d\alpha \frac{e^{ip\alpha -im\theta(\alpha)}}{ip} + 
\frac{1}{ip} \oint d\alpha
\frac{\theta''(\alpha)}{im(\theta'(\alpha))^2} e^{ip\alpha -
  im\theta(\alpha)}  
\end{equation}
To estimate the second term we notice that
$|e^{ip\alpha-im\theta(\alpha)}| \leq 1$. Then taking $\theta(\alpha)$
as given by Eq. \ref{thetaapprox} then we find  
\begin{equation}
\begin{array}{ll}
\left| 
\oint \frac{\theta''(\alpha)}{im (\theta'(\alpha))^2} e^{ip\alpha
  -im\theta(\alpha)}  
\right| 
& \lesssim 
\left| 
\frac{q}{mc_t} \oint 
\frac{\frac{\alpha}{2}}{\sqrt{ \frac{\alpha^2}{2} + \frac{\epsilon^2}{2a^2
} }} d\alpha 
\right| 
\\
& =
\frac{q}{mc_t} \sqrt{2} 
\left( \frac{\epsilon}{a} \right) 
\left| \ln \left( \frac{2\pi a}{\epsilon} \right) 
\right| 
\\
& \sim 
\frac{ 
\left( \frac{\epsilon}{a} \right) 
\left| \ln \left( \frac{\epsilon}{a} \right) \right| }{n}
\end{array}
\end{equation}
where the integral is easily obtained by substituting
$\frac{a\alpha}{\epsilon} = \mbox{Sinh}(u)$.  
Hence for $\epsilon /a \sim 1$ the term is of order $1/n$ and may be
neglected, and as $\epsilon /a \rightarrow 0$ the term also tends to
zero, and hence may be neglected.  
Thus in the high-n limit we get 
\begin{equation}\label{dW3}
\delta W_V = \pi \Delta \left( \frac{m}{nq} \right)^2  \xi_m^* 
\left(\frac{r_a^2I^2}{R^2B^2} \right) 
\sum_{p\neq 0} a_p \frac{i|p|(|p|+1)}{pr_a^2} 
\oint d\alpha \mbox{ } e^{-im\theta(\alpha) +ip\alpha} 
\end{equation}

Taking the complex conjugate of Eq. \ref{ap} and rearranging, we get 
\begin{equation}
\oint d\alpha e^{-im\theta(\alpha)+ip\alpha} = i a_p^* \frac{|p|}{p}
\frac{R}{\xi_m^*} \frac{2\pi}{\Delta}  
\end{equation}
which upon substitution into Eq. \ref{dW3}, gives 
\begin{equation}\label{dW4}
\delta W_V = 2\pi^2 R \left( \frac{m}{nq} \right)^2  
\left(\frac{I^2}{R^2B^2} \right) 
\sum_{p\neq 0} \left( \left| p \right| + 1 \right) \left| a_p\right|^2 
\end{equation}
For the high $m$,$n$ limit considered here, we expect the $|a_p|^2$
coefficients to be largest for $p\sim m\sim nq$, and hence in the high
$m$, $n$ limit we expect  
\begin{equation}\label{dW5}
\begin{array}{l}
\delta W_V = 2\pi^2 R 
\sum_{p\neq 0} \left| p \right|  \left| a_p\right|^2 
\end{array}
\end{equation}
where we have also taken $I^2/(R^2B^2) \simeq 1$. 
Hence we have re-obtained Eq. \ref{FdWV} from the high-$n$ expression
given in part (I). 
This gives us confidence in the reliability of the calculations. 
In addition $[| n_z . \nabla_z (n_z.B_z)|]$ may be estimated by
approximating the plasma as a vacuum and solving Laplace's equation to
approximate and obtain the perturbed field both inside and outside the
plasma respectively, and correctly matching the fields at the
plasma-vacuum boundary.  
Then we find that $[| n_z . \nabla_z
  (n_z.B_z)|]=2n_z.\nabla_z(n_z.B_z)$. Hence the above calculation may
be used to infer that the term $-\pi |\xi_m|^2 \Delta^2 \Delta'$
appearing in Eq. 70 of the first part to this paper, is equal to
$2\delta W_V$, where   
\begin{equation}
\Delta' \equiv \left[ \left| \frac{1}{2\pi} 
\oint dl RB_p \frac{I^2}{R^2B^2} \frac{\frac{\partial}{\partial \psi}
  \nabla \psi .\vec{B}_1 } {\nabla \psi . \vec{B}_1} \right| \right]  
\end{equation}
Hence to evaluate $\Delta'$, we need solely evaluate $\delta W_V$.

\section{Directly calculating $\Delta'$}\label{secDelta'}

Here we show how to calculate $\Delta'$ directly, using the same
assumptions as in Section \ref{recalc}. 
For simplicity in all that follows we will take $I^2/R^2 B^2\simeq 1$,
and firstly use Eq. \ref{EPSBC}, that    
$\left. \nabla \psi . \vec{B}_1^V \right|_{edge}= \left. \nabla \psi
.\vec{B}_1 \right|_{edge}$ to write  
\begin{equation} 
\Delta' = \left[ \left| \frac{1}{2\pi} 
\oint dl RB_p \frac{I^2}{R^2B^2} \frac{\frac{\partial}{\partial \psi}
  \nabla \psi .\vec{B}_1 } {\nabla \psi . \vec{B}_1} \right| \right]  
= \frac{1}{2\pi} 
\oint dl RB_p \frac{ \left[ \left| \frac{\partial}{\partial \psi}
    \nabla \psi .\vec{B}_1 \right| \right] }{\nabla \psi .\vec{B}_1}  
\end{equation}
Next we use Eqs. \ref{transforms}, to obtain 
\begin{equation}
\Delta' = \frac{1}{2\pi}
\oint r_a d\alpha \frac{1}{n_z . B_z} \left[ \left| n_z . \nabla
  \left( n_z . B_z \right) \right| \right]  
\end{equation}
Making the usual approximation that treats the perturbed field near
the plasma's edge as behaving the same as in a vacuum, and also using
Eq. \ref{Vvac}, gives  
\begin{equation}
\left[ \left| n_z . \nabla_z \left( n_z . B_z \right) \right| \right] 
= 2 \left. \frac{\partial^2 V}{\partial r^2} \right|_{r=r_a} 
= 2 \sum_{p \neq 0} e^{ip\alpha -in\phi} \frac{a_p}{r_a^2} |p| \left(
|p| + 1 \right)  
\end{equation}
We also have
\begin{equation}
\begin{array}{ll}
n_z . B_z  = \left. \frac{\partial V}{\partial r} \right|_{r=r_a} 
&= - \sum_{p\neq 0} e^{ip\alpha-in\phi} a_p \frac{|p|}{r_a} 
\\
&= \Delta \frac{\xi_m}{R} \frac{e^{-in\phi}}{r_a} \sum_{p\neq 0}
e^{ip\alpha} \frac{1}{2\pi} \oint im\theta'(\beta)
e^{im\theta(\beta)-ip\beta} d\beta  
\\
&=  \Delta \frac{\xi_m}{R} \frac{e^{-in\phi}}{r_a} 
\theta'(\alpha) e^{im\theta(\alpha)} 
\end{array}
\end{equation}
where we used Eq. \ref{theta1} that implies $\theta'(\alpha) =
\frac{Ir_a}{qR^2} \frac{|w'|^2}{|B_{pz}|}$, Eq. \ref{ap1}, and that  
$\theta'(\alpha) e^{im\theta(\alpha)} = \sum_{p\neq0} 
e^{ip\alpha} \frac{1}{2\pi} \oint \theta'(\beta) e^{im\theta(\beta)
  -ip\beta} d\beta$.  
Note that the above equation can be obtained more directly from
Eq. \ref{nzBC}.  

Using the above results, after some cancellations we obtain 
\begin{equation}
\begin{array}{ll}
\Delta' &= \frac{2}{2\pi} \oint d\alpha \frac{R}{\xi_m \Delta}
\frac{e^{-im\theta(\alpha)}}{im\theta'(\alpha)} \sum_{p\neq 0}
e^{ip\alpha} a_p |p| \left( |p| + 1 \right)  
\\
&= \frac{1}{\pi} \frac{R}{\xi_m \Delta} \sum_{p\neq 0} a_p |p|^2 \oint
d\alpha \frac{e^{-im\theta(\alpha)+ip\alpha} }{im\theta'(\alpha)}  
+ \frac{1}{\pi} \frac{R}{\xi_m \Delta} \sum_{p\neq 0} a_p |p| \oint
d\alpha \frac{e^{-im\theta(\alpha)+ip\alpha} }{im\theta'(\alpha)}  
\end{array}
\end{equation}
Then using the result Eq. \ref{keyeq} of Section \ref{recalc}, that 
\begin{equation}
\begin{array}{ll}
\oint d\alpha \frac{e^{-im\theta(\alpha)}}{im\theta'(\alpha)} e^{ip\alpha} 
&= \frac{1}{ip} \oint e^{-im\theta + ip\alpha} + \mbox{O} \left( 
\frac{\left( \frac{\epsilon}{a} \right) \left| \ln \left(
  \frac{\epsilon}{a} \right) \right| }{n}  
\right)  
\end{array}
\end{equation}
we obtain the first term as
\begin{equation}
\begin{array}{ll}
\frac{R}{\pi\xi_m \Delta} \sum_{p\neq 0} a_p |p|^2 \oint d\alpha
\frac{e^{-im\theta(\alpha)+ip\alpha} }{im\theta '(\alpha)}  
&= \frac{2}{|\xi_m |^2} 
\frac{R^2}{\Delta^2} 
\sum_{p \neq 0} a_p |p| 
\frac{(-ip)}{|p|} \Delta \frac{\xi_m^*}{R} \frac{1}{2\pi} 
\oint e^{ip\alpha -im\theta} 
\\
&= \frac{2 R^2}{|\xi_m|^2 \Delta^2} (-1) \sum_{p\neq 0} |p| |a_p|^2 
\end{array}
\end{equation}
Using Eq. \ref{ap} for $a_p$, and the same approximations as above,
the second term gives  
\begin{equation}
\begin{array}{ll}
\frac{R}{\pi \xi_m \Delta} \sum_{p\neq 0} |p| a_p \oint d\alpha
\frac{e^{-im\theta(\alpha) +ip\alpha}}{im\theta'(\alpha)}  
&= \frac{R}{\pi \xi_m \Delta} \sum_{p\neq 0} |p| a_p \frac{1}{ip} \oint
e^{-im\theta +ip\alpha} + \mbox{O}\left(\frac{1}{nq} \right)  
\\
&= -\frac{1}{\pi} \sum_{p\neq 0} \oint e^{im\theta(\beta)-ip\beta}
d\beta \frac{1}{2\pi} \oint e^{im\theta(\alpha)-ip\alpha} d\alpha  
\\
&= -\frac{1}{\pi} \oint d\alpha e^{-im\theta(\alpha)} \sum_{p\neq 0}
e^{ip\alpha} \frac{1}{2\pi} \oint e^{im\theta(\beta)-ip\beta} d\beta  
\\
&= -\frac{1}{\pi} \oint e^{-im\theta(\alpha)} e^{im\theta(\alpha)} d\alpha 
= - 2 
\end{array}
\end{equation}
Hence using all of the above, and Eq. \ref{FdWV} for $\delta W_V$, 
\begin{equation}
\Delta' = -2 \left\{ \frac{R^2}{|\xi_m|^2 \Delta^2} \sum_{p\neq 0}
|p| |a_p|^2 + 1 \right\} 
= -2 \left( \frac{\delta W_V}{2\pi^2\frac{|\xi_m|^2}{R} \Delta^2}
\right) \left\{ 1 + \mbox{O} \left( \frac{1}{\delta W_V} \right)
\right\}  
\end{equation}
(Later we will find that $\sum_{p\neq0} |p||a_p|^2 = m \frac{|\xi_m|^2 
 \Delta^2}{R^2}$ and hence that $\Delta' \simeq - 2 m$.)

\section{Evaluating the sum}\label{Sum1}

We have found that the vacuum energy $\delta W_V$ and $\Delta'$ are
both determined from $\sum_{p=-\infty}^{\infty} |p||a_p|^2$, 
that we may in principle evaluate using our analytical expression for
$a_p$. 
We do that here. 

Firstly note that if $|p|$ is replaced with $p$ in Eq. \ref{FdWV},
then we may easily resum the series, because 
\begin{equation}\label{exactsum}
\begin{array}{ll}
\sum_{p=-\infty}^{\infty} p |a_p|^2 
&= \Delta^2 \frac{|\xi_m|^2}{R^2} \sum_{p=-\infty}^{\infty} 
p \frac{1}{2\pi} \oint d\alpha \mbox{ } e^{im\theta (\alpha)-ip\alpha} 
\frac{1}{2\pi} 
\oint d\beta \mbox{ } e^{-im\theta (\beta) +ip\beta} 
\\
&= \frac{\Delta^2}{2\pi}  \frac{|\xi_m|^2}{R^2} 
\oint d\beta \mbox{ } e^{-im\theta (\beta)} \sum_{p=-\infty}^{\infty}
e^{ip\beta}  
\frac{1}{2\pi} \oint d\alpha \mbox{ } p \mbox{ } e^{im\theta(\alpha)-ip\alpha} 
\\
&= \frac{\Delta^2}{2\pi}  \frac{|\xi_m|^2}{R^2} 
\oint d\beta \mbox{ } e^{-im\theta (\beta)} \mbox{ } \sum_{p=-\infty}^{\infty} 
e^{ip\beta} 
\frac{1}{2\pi} \oint d\alpha \mbox{ } m\theta'(\alpha)
e^{im\theta(\alpha)-ip\alpha}  
\\
&= \frac{\Delta^2}{2\pi}  \frac{|\xi_m|^2}{R^2} m
\oint d\beta \mbox{ } e^{-im\theta (\beta)} \theta'(\beta) e^{im\theta(\beta)} 
= \Delta^2 \frac{|\xi_m|^2}{R^2} m
\end{array}
\end{equation}
where in going from lines $3$ to $4$ we integrated by parts, and in
going from lines $4$ to $5$ we note that $\theta'(\beta)
e^{im\theta(\beta)}= \sum_{p\neq 0} e^{ip\beta} 
\frac{1}{2\pi} \oint \theta'(\alpha)
e^{im\theta(\alpha)} e^{-ip\alpha} d\alpha$.  
We might expect the values of the coefficients to be peaked for values
of $p\sim m  \sim nq \gg 1$, and so it is likely that
$\sum_{p=-\infty}^{\infty} |p||a_p|^2 \simeq
\sum_{p=1}^{\infty}p|a_p|^2 \simeq \sum_{p=-\infty}^{\infty}p|a_p|^2$.
This has been confirmed by calculating the sums using a saddle point 
approximation. 
(The details of the calculation are too long to be included here)

It is instructive to recalculate the result using a simple 
model for $\theta (\alpha)$, that encapsulates the fact that as we
approach the 
separatrix (with $q\rightarrow \infty$ and the local field line pitch 
$\nu$ becoming increasingly peaked near the X-point), the function
$\theta(\alpha)$ becomes increasingly similar to a step function.  
In this simple model we take $q \sim \frac{1}{\delta}$, $\delta \ll
1$, and $\nu \sim q \sim \frac{1}{\delta}$ when $\alpha \in
(-\delta,\delta)$, this leads to a  
very simple model for $\theta(\alpha)$, with 
\begin{equation}
\theta(\alpha)=
\left\{ 
\begin{array}{ll}
0 & \alpha \in (-\pi, -\delta)
\\
\left( \frac{\alpha+\delta}{2\delta} \right) 2\pi & \alpha \in
(-\delta, \delta ) 
\\
2\pi &\alpha \in (\delta, \pi)
\end{array}
\right.
\end{equation}
So that 
\begin{equation}
e^{im\theta(\alpha)}=
\left\{ 
\begin{array}{ll}
1 & \alpha \in (-\pi, -\delta)
\\
e^{im\left( \frac{\alpha+\delta}{2\delta} \right) 2\pi} & \alpha \in
(-\delta, \delta ) 
\\
1 &\alpha \in (\delta, \pi)
\end{array}
\right.
\end{equation}
Because $\theta(\alpha)$ is piecewise linear, it is easy to evaluate
$\oint d\alpha e^{im\theta(\alpha) - ip\alpha}$, that gives  
\begin{equation}
a_p = -i \frac{p}{|p|} \Delta \frac{\xi_m}{R} m
\frac{\sin(p\delta)}{p(p\delta-m\pi)}  
\end{equation}
Hence
\begin{equation}
\begin{array}{ll}
\delta W_V &= 2\pi^2 R \sum_{p\neq 0} |p| |a_p|^2 
\\
&= 2\pi^2 R \frac{|\xi_m|^2}{R^2} \Delta^2 m^2 \sum_{p\neq 0}
\frac{\sin^2(p\delta)}{|p|(p\delta-m\pi)^2}  
\\
& \simeq \frac{2\pi^2}{R} |\xi_m|^2 \Delta^2 m^2 \left(
\int_1^{\infty} \frac{\sin^2(p\delta)}{|p|(p\delta-m\pi)^2} dp +
\int_{1}^{\infty} \frac{\sin^2(p\delta)}{|p|(p\delta+m\pi)^2} dp
\right)  
\\ 
& \rightarrow  \frac{2\pi^2}{R} |\xi_m|^2 \Delta^2 m^2 \frac{1}{m} 
 \mbox{  for } m \gg 1 
\end{array}
\end{equation}
the same as was obtained previously by the saddle point
approximation.  
To obtain this result we used 
\begin{equation}
\sum_{p=1}^{\infty} \frac{\sin^2(p\delta)}{|p|(p\delta-m\pi)^2} \simeq 
\int_1^{\infty} \frac{\sin^2(p\delta)}{|p|(p\delta - m\pi)^2} dp = 
\frac{1}{m} + \mbox{O} \left( \frac{\ln (\delta)}{m^2} \right) = 
\frac{1}{m} + \mbox{O} \left( \frac{\delta \ln (\delta)}{m} \right)
\end{equation}
and 
\begin{equation}
\sum_{p=-\infty}^{-1} \frac{\sin^2(p\delta)}{|p|(p\delta-m\pi)^2} \simeq 
\int_1^{\infty} \frac{\sin^2(p\delta)}{|p|(p\delta + m\pi)^2} dp = 
\mbox{O} \left( \frac{\ln (\delta)}{m^2} \right)=  
\mbox{O} \left( \frac{\delta \ln (\delta)}{m} \right)
\end{equation}
where we also used $m\sim nq \sim \frac{n}{\delta}$. 
Notice that the integrals do not diverge at $p\delta-m\pi=0$, because
$\sin(p\delta)=\sin(p\delta-m\pi)=0$ for $p\delta-m\pi=0$, this would
not be the case if $m$ were not an integer.
Hence not only do we find agreement with the calculation using the
saddle point approximation, but we again find that  
\begin{equation}
\sum_{p=-\infty}^{\infty} |p| |a_p|^2 \rightarrow
\sum_{p=-\infty}^{\infty} p|a_p|^2 
\mbox{   as   } m\sim nq \rightarrow \infty 
\end{equation}
Therefore both methods suggest that provided $m\gg 1$ and $n\gg1$, then 
\begin{equation}
\sum_{p=-\infty}^{\infty} |p| |a_p|^2 \rightarrow
\sum_{p=-\infty}^{\infty} p|a_p|^2 = \Delta^2 \frac{|\xi_m|^2}{R^2} m  
\end{equation}
In addition notice 
that for $m\sim nq \gg 1$ the result of neither approximation methods
involve $\delta$ (or $\epsilon$), suggesting that the result may be
generic and independent of the detailed form of $\theta(\alpha)$.

Returning to the calculation of $\Delta'$, Sections \ref{recalc} and
\ref{secDelta'} showed that at leading order  
\begin{equation}
\Delta' = -2  \left( \frac{\delta W_V}{2\pi^2\frac{|\xi
    |^2}{R}\Delta^2} \right) 
\end{equation}
and that using the above results gives 
\begin{equation}
\Delta' = - 2 m 
\end{equation}
In addition, the work described above suggests that the result is generic 
for perturbations with $n\gg 1$, regardless of whether the plasma
cross-section is circular, or shaped with a separatrix boundary that
contains an X-point.

\section{Discussion}\label{Conc1}

\subsection{Scope \& Purpose of the Calculation}

In the first part to this paper we started from the simplest model
used to study Peeling modes, that considers Peeling modes in a
cylindrical plasma at marginal stability, then  generalised it to a
toroidal plasma.  
According to the model, the energy principle's $\delta W$ 
is determined by the value of $\Delta'$, that is a normalised measure
of the jump in the gradient of the normal component of the perturbed
magnetic field.  
In this second part we have restricted ourselves to systems for which
the vacuum magnetic field may be treated as being approximately two
dimensional, as is the case for a sufficiently large aspect ratio
Tokamak.  
This allows us to use a conformal transformation in our calculations,
and at high toroidal mode number we have obtained analytic expressions
for the vacuum energy and $\Delta'$, whenever the plasma is perturbed
by a radial displacement consisting of a Fourier mode in straight
field line co-ordinates.  
These expressions remain valid for a plasma cross-section that
approximates a separatrix with an X-point, and appear to be generic,
independent of the exact form for $\theta(\alpha)$.  
Because it is possible to do this analytically, there is the
possibility of making similar analytic progress with other linear
plasma instabilities whose plasma equilibria have a separatrix with
an X-point.  
Such calculations can provide physical understanding and useful tests
during the development of codes to study the stability of more general
geometry Tokamak plasmas, either giving confidence in a code or
indicating its limitations.

\subsection{(In)Stability of the Ideal MHD Peeling Mode?}

According to the model developed in part (I) and our calculation here
of $\Delta' = -2m$, we can now examine the model's predictions.  
According to the model developed in part (I), for the trial function
used by Laval et al\cite{Laval}, stability is determined by the sign
of  
\begin{equation}
\delta W = -2\pi^2 \frac{|\xi_m|^2}{R} \Delta
\left[ \Delta \Delta' + \hat{J} \right] 
\end{equation}
with 
\begin{equation}
\Delta=\frac{m-nq}{nq}
\end{equation}
\begin{equation}
\hat{J}= \frac{1}{2\pi} \oint dl \frac{I}{RB_p}
\frac{\vec{J}.\vec{B}}{B^2}
\end{equation}
\begin{equation}
\hat{\Delta}'= \left[ \left|
 \frac{1}{2\pi} \oint dl R B_p \frac{I^2}{R^2B^2}
\frac{\frac{\partial}{\partial \psi} \nabla \psi
    .\vec{B}_1}{\nabla \psi .\vec{B}_1} \right| \right]
\end{equation}
and as mentioned in Section \ref{D}, $R=R_0$ is constant for the large
aspect ratio limit considered here.  
$\delta W$ is minimised with respect to $\Delta$ (or equivalently, a
particular choice of toroidal mode number), finding
$\Delta=-\hat{J}/(2\Delta')$. 
Then using our result of $\Delta' = -2m$ we get 
\begin{equation}\label{minimiseddW}
\delta W = -\left(\frac{\pi}{2}\right)^2 \frac{|\xi|^2}{R} 
\left( \frac{\hat{J}^2}{m} \right) 
\end{equation}
When checking the dimensions of $\delta W$ it is essential to remember
that $\xi_m = (\nabla \psi .\vec{\xi})_m \sim  rRB_p $, then because
$B^2 \sim p$ is an energy per unit volume, $\delta W \sim r^2 RB_p^2$
and hence has units of energy. 
For the process of minimisation $n$ was treated as a continuous
variable (that for $m\sim nq \gg 1$ is a reasonable approximation).  
Now we consider two cases in turn, firstly $\nabla \phi . \vec{J}=0$,
for which  
\begin{equation}
\hat{J} = \frac{1}{2\pi} \oint dl \frac{I}{RB_p} \frac{-I'B_p^2}{B^2} 
= \frac{1}{2\pi} \oint dl B_p \frac{-II'}{RB^2} 
\sim 1 
\end{equation}
So that although $\delta W <0$ for all $m$, because $m\sim nq
\rightarrow \infty$ then $\delta W \sim \frac{1}{m} \rightarrow 0$.  
For the case of $\nabla \phi . \vec{J} \neq 0$ however, 
\begin{equation}\label{Jhatphi}
\hat{J} = \frac{1}{2\pi} \oint dl \frac{I}{RB_p} \frac{\vec{J}.\vec{B}}{B^2} 
= \frac{1}{2\pi} \oint \frac{dl}{B_p} \frac{I}{R^2} R
\frac{\vec{J}.\vec{B}}{B^2}  
\simeq \left\langle R \frac{\vec{J}.\vec{B}}{B^2} \right\rangle
\frac{1}{2\pi} \oint \frac{dl}{B_p} \frac{I}{R^2}  
= q \left\langle R \frac{\vec{J}.\vec{B}}{B^2} \right\rangle
\end{equation}
where $\left\langle \right\rangle$ denotes the poloidal average, and 
$q= \frac{1}{2\pi} \oint \frac{I}{R^2} \frac{dl}{B_p}$ is the safety
factor\cite{Laval}.
The divergence in $\frac{\nu}{B_p^2}$ at the X-point will mean that
the poloidal location of the X-point will affect the value of
$\hat{J}$ (as a function of $q$), this is also the case for Mercier
stability and is discussed by Webster\cite{Webster08}. 
Then using Eqs. \ref{minimiseddW}, \ref{Jhatphi}, and $m\sim nq$ we have  
\begin{equation}
\begin{array}{ll}
\delta W &\sim  - \left( \frac{\pi}{2} \right)^2 \frac{|\xi_m|^2}{R}
\frac{q}{n}  
\langle R \frac{\vec{J}.\vec{B}}{B^2} \rangle^2 < 0
\end{array}
\end{equation}
Therefore if $\delta W<0$ is taken to indicate instability, the
result would indicate that the Peeling mode remains unstable near a
separatrix.  
However as observed in the first part of this paper the growth rate
$\gamma$ has  
$\gamma^2 = - \delta W /\int \vec{dr} \rho_0 |\xi |^2$, with $\int
  \vec{dr} \rho_0 |\xi |^2$ diverging at a rate proportional to
  $q'(\psi)$.  
This gives $\ln(\gamma)= -\frac{1}{2} \ln (q'/q)$ for the limit of a
separatrix with $q$ and $q'$ tending to infinity, so that although
$\delta W$ is non-zero and negative, the mode will be marginally
stable.  
This is similar to the calculation of the Mercier coefficient by
Webster\cite{Webster08}, where $D_M$ is found from the ratio of two 
diverging quantities, with $D_M\sim q/q'\rightarrow 0$ as the
separatrix is approached. 
Note that these results for $\Delta'$ and the growth rate appear to be
generic and independent of the detailed forms of $\theta(\alpha)$ or
the radial structure of the mode.

\subsection{Is the Mode Physically Acceptable?}

Because we require a poloidal mode number $m\sim nq$, then near the
separatrix where $q \rightarrow \infty$ we also require $m\rightarrow
\infty$. 
This raises the question: Is the mode physically acceptable?
To answer this we reconsider the trial function, that has $\xi \sim
e^{im\theta(\alpha)}$, with 
$\theta(\alpha)=\frac{1}{q}\int^{\alpha} \nu(\alpha) d\alpha$. 
When $m\sim nq$ the trial function becomes
 $\xi \sim e^{in\int^{\alpha} \nu d\alpha}$, from which we can see
that because $\nu$ is of order one and well behaved everywhere except
near the x-point (where it diverges to infinity), then so also is the
mode's structure.  
As discussed in part (I),
the divergence in $m\sim nq \rightarrow \infty$ is
only manifested in close proximity to the x-point where the divergence
in $\nu$  causes the mode to oscillate increasingly
rapidly as the x-point is approached.  
Elsewhere $\nu$ is typically of order $1$, and the mode structure is
like that for a finite 
mode number, oscillating at a modest rate of order $n\nu \ll nq$, and 
only weakly affected by the proximity of the flux surface to the
separatrix.  
Hence the mode has a simple structure everywhere except for a region
close to the x-point where it oscillates so rapidly that MHD would no
longer be applicable.  
The resulting mode structure is consistent with
observations of ELMs\cite{Kirk}, that show filamentary structures
that follow the magnetic field lines, and whose poloidal structure
near the X-point is difficult to determine.

\subsection{Previous Analytical Work}\label{analytical1} 

As mentioned at the outset, Laval et al\cite{Laval}  
considered a trial function consisting of a single Fourier mode in a
straight field line co-ordinate, that is resonant at a rational
surface in the vacuum just outside the plasma's surface. 
For that trial function, they found that  
for a positive non-zero current at the plasma edge, $\delta
W <0$, and suggested therefore that the Peeling mode would be
unstable for a non-zero positive current at the plasma's edge.  
On the basis of the sign of $\delta W$ our study also finds this, but
our study also suggests that the growth rate will asymptote to
zero as the 
outermost flux surface approximates a separatrix, so that the mode will
be marginally stable.

Lortz\cite{Lortz} also considered Peeling mode stability, in toroidal
plasmas with shaped cross-sections, using a systematic calculation
with a trial function whose resonant surface is inside the plasma. 
An advantage of the calculation by Lortz\cite{Lortz}, is that the
radial structure of the mode is considered. 
An unfortunate complication for this discussion is that in the
ordering scheme of Lortz\cite{Lortz}, the vacuum energy can be
neglected. 
This was not the case for our calculation in Part (I)\cite{part1}, or
of Laval et al\cite{Laval}. 
Nonetheless, we will consider the predictions of this calculation in
the limit where the outermost flux surface approximates a separatrix.

Connor et al\cite{Connor} review the calculation of Lortz\cite{Lortz},
and use it to consider trial functions with resonant surfaces both
inside and outside the plasma. 
They find that stability of the Peeling mode
requires\cite{Connor}  
\begin{equation}\label{HastieStab}
1-4D_M > \left( 2 \frac{S}{P} - 1 \right)^2 
\end{equation}
where the Mercier co-efficient $D_M \equiv -Q/P$, and $P$, $Q$, $S$
are defined as,  
\begin{equation}
\begin{array}{c}
P = 2\pi (q')^2 \left[ 
\oint \frac{J_{\chi}B^2}{R^2B_p^2}  d\chi 
\right]^{-1} 
\\
Q = \frac{p'}{2\pi} 
\oint \frac{\partial J_{\chi}}{\partial \psi} d\chi 
- \frac{(p')^2}{2\pi} 
\oint \frac{J_{\chi}}{B_p^2} d\chi 
+ Ip' \oint \frac{J_{\chi}}{R^2B_p^2} d\chi 
\left[ 
\oint \frac{J_{\chi}B^2}{R^2B_p^2} d\chi 
\right]^{-1}
\times \left[
\frac{Ip'}{2\pi} \oint \frac{J_{\chi}B^2}{R^2B_p^2} - q' 
\right]
\\
S=P + q' \oint \frac{j_{\parallel} B}{R^2B_p^2} J_{\chi} d\chi \left[
\oint \frac{J_{\chi} B^2}{R^2 B_p^2} d\chi 
\right]^{-1}
\end{array}
\end{equation}
Substituting $D_M \equiv - Q/P$ into Eq. \ref{HastieStab}, gives the
stability requirement  
\begin{equation}\label{SCon1}
\frac{Q}{P} - \left( \frac{S}{P} \right)^2 + \frac{S}{P} > 0
\end{equation}
Substituting for $P$, $Q$, and $S$, allows Eq. \ref{SCon1} to be
simplified to 
\begin{equation}\label{SCon2}
p' \left[ 
\frac{\partial }{\partial \psi} 
\frac{1}{2\pi} \oint J_{\chi} d\chi 
- p' \frac{1}{2\pi} 
\oint \frac{J_{\chi}}{B_p^2} d\chi \right] 
+I' \left[ 
q' - 
2p' \frac{1}{2\pi} \oint \frac{\nu}{B_p^2} d\chi  
- I' \frac{1}{2\pi} \oint \frac{J_{\chi}B^2}{R^2B_p^2} 
\right]>0
\end{equation}
This may be simplified further by noting that because
$\nu=IJ_{\chi}/R^2$, and in a large aspect ratio ordering where 
$R$ is taken as approximately constant,  
\begin{equation}
\frac{\partial}{\partial \psi} \frac{1}{2\pi} \oint J_{\chi} d\chi 
= \frac{\partial}{\partial \psi} \frac{1}{2\pi} \oint \frac{\nu R^2}{I} d\chi
= \frac{R^2}{I} \frac{1}{2\pi} \oint \frac{\partial \nu}{\partial
  \psi} d\chi  
- R^2 \frac{I'}{I^2} \frac{1}{2\pi} \oint \nu d\chi 
\end{equation}
and 
\begin{equation}
q' = \frac{1}{2\pi} \oint \frac{\partial \nu}{\partial \psi} d\chi 
\end{equation}
The Grad-Shafranov equation in $\psi$, $\chi$, $\phi$
co-ordinates, has 
\begin{equation}\label{dnudpsi}
\frac{\partial \nu}{\partial \psi} = \frac{\nu}{B_p^2} \left\{ 
- \frac{\partial}{\partial \psi} \left( p+ B^2 \right) +
\frac{R^2B^2}{I} \frac{\partial}{\partial \psi} \left( \frac{I}{R^2}
\right)  
\right\}
\end{equation}
Therefore if $R$ is taken as approximately constant (as would be the
case either 
in a large aspect ratio limit or if we are sufficiently close to the
separatrix that the integral is dominated by the divergence at the
X-point and $R$ may be approximated by its value $R=R_X$ there), then
using Eq. \ref{dnudpsi},  Eq. \ref{SCon2} simplifies to a condition
for stability of   
\begin{equation}
\begin{array}{ll}
0&< \frac{1}{2\pi} \oint \frac{\nu}{B_p^2} \frac{I}{R^2} d\chi \left\{ 
\left( -p' - \frac{II'}{R^2} \right) \frac{\partial}{\partial \psi}
\left( 2p + B^2 \right) \right\}   
\\&=
\left( \nabla \phi . \vec{J} \right) I \frac{1}{2\pi} 
\oint \frac{\nu}{R^2B_p^2} 
\frac{\partial}{\partial \psi} \left( 2p + B^2 \right) d\chi 
\end{array}
\end{equation}
This may be simplified further still, to 
\begin{equation}
0< \frac{1}{2\pi} 
\oint \frac{\nu}{B_p^2} \frac{I}{R^2}
\left( - \nabla \phi . \vec{J} \right) 
\left[  2 \left( \nabla \phi . \vec{J} \right) 
- \frac{\partial B_p^2}{\partial \psi} \right]d\chi 
\end{equation}
Because the integrals are dominated by the divergence of $\nu/B_p^2$
at the X-point and  $\frac{\partial B_p^2}{\partial \psi} <0$
at the X-point, then based on the formulation of
Lortz\cite{Lortz,Connor}, then  
provided $\nabla \phi . \vec{J}>0$ the negative
expression clearly indicates instability to the Peeling mode. 
However, if we allow $\nabla \phi . \vec{J}$ to be negative, then the
formulation of Lortz\cite{Lortz,Connor} also  suggests that 
stability is possible provided 
\begin{equation}
0 < 
\frac{1}{2\pi} \oint \frac{\nu}{B_p^2} \left[ -2\left| \nabla \phi
  . \vec{J} \right| - \frac{\partial B_p^2}{\partial \psi} \right]
  d\chi 
\end{equation}
Therefore in principle there is a range of negative current values at
the plasma edge for which the Peeling mode is stable. 
The appendix calculates $\partial B_p^2/\partial \psi$ near the
X-point for a standard and a ``snowflake''\cite{Ryutov} divertor. 
Interestingly, whereas a conventional X-point has a range of negative
current values for which the Peeling mode is stable, in the limit of
an exact ``snowflake'' X-point (with flux surfaces meeting at an angle
of $\pi/3$), the range of values of negative current for which the
Peeling mode is stable, tends to zero. 
Whether this observation will have consequences for the plasma
behaviour in a ``snowflake'' X-point geometry remains to be seen, but
it is a qualitative difference between a conventional X-point and that
produced with a ``snowflake'' divertor.


As mentioned previously, the calculation 
also considered the radial structure of the mode, with $\xi \sim
x^{\lambda_{\pm}}$, $x$ a radial co-ordinate, and  
\begin{equation}
\lambda_{\pm} = - \frac{1}{2} \pm \sqrt{ \frac{1}{4} + \frac{Q}{P} } 
\end{equation} 
As we approach the separatrix, Webster\cite{Webster08} shows that 
$D_M=-\frac{Q}{P} \rightarrow 0$, giving 
\begin{equation}
\begin{array}{l}
\lambda_{+} \simeq \frac{Q}{P} \rightarrow 0
\\
\lambda_{-} \simeq -1 
\end{array}
\end{equation}
and mode structures of $\xi_{-} \sim \frac{1}{x}$ and $\xi_{+} \sim x^{Q/P}$. 
For the perturbations to satisfy the boundary condition of a mode
amplitude that tends to zero in the plasma, this requires us to use
$\xi_{-}$ for resonances outside the plasma (the ``external'' Peeling
mode), and $\xi_{+}$ for resonances inside the plasma (the
``internal'' Peeling mode).
It should be noted that a 
potential problem with the analysis of Lortz et al\cite{Lortz} when
applied to Peeling modes, that the mode is taken to be sufficiently
localised that the equilibrium quantities (that include $q$ and $q'$),
are approximately constant.  
This is almost certainly not the case near a separatrix.

\subsection{Summary}

We have started from a simple model for the Peeling mode, at
marginal stability in cylindrical geometry, and in Part (I) of
this paper generalised it to toroidal Tokamak geometry.  
A conclusion of Part (I) is that Peeling mode
stability is determined by the value of $\Delta'$, a normalised
measure of the discontinuity in the gradient of the normal component
of the perturbed magnetic field at the plasma-vacuum boundary.  
Therefore this paper evaluated $\Delta'$ in a large aspect
ratio Tokamak geometry with a separatrix and X-point, but in such a
way that the effect of the X-point is captured exactly, without
encountering the usual discretisation errors present in most numerical
methods.  
This was possible by generalising the method of conformal
transformations beyond textbook presentations, that require a boundary
condition of either the function or its normal derivative to be zero. 
Here we observe that even if the field's normal derivative is
non-zero at the boundary, it is still possible to use the conformal
transformation method.  
In this case instead of obtaining an exact analytic solution (as would
be the case if its normal derivative were zero on the boundary), the 
2-dimensional problem is reduced to a 1-dimensional problem that may
subsequently be solved exactly or approximated.  
The approach avoids the errors that may arise due to the
discretisation of space near an X-point, that are necessarily 
present in most numerical methods. 
This paper also calculated analytical expressions 
for physically realistic examples of the equilibrium vacuum magnetic
field, and the straight field line angle at the plasma-vacuum
boundary.  
These and other results are likely to find opportunities for
application elsewhere.  

It is found that a radial plasma perturbation
consisting of a single Fourier mode in straight field line
co-ordinates with a high toroidal mode number $n$, in a plasma
equilibrium with a separatrix and an x-point, will produce the same
change in the vacuum energy as the equivalent perturbation in a
cylindrical equilibrium, with $\delta W_V = 2\pi^2 \frac{|\xi_m|^2}{R}
\Delta^2 m$.  
It also results in the same value for $\Delta'$, with $\Delta'=-2m$,
where $m$ is the poloidal mode number.  
Despite our trial function requiring $m\sim nq \rightarrow \infty$, we
observe that the trial function has $\xi \sim e^{in \int^{\chi} \nu
  d\chi}$ that is physically well  behaved for all but a highly
localised region near the X-point where MHD will fail to apply.  
Therefore we believe the trial function is physically acceptable even
for a separatrix  boundary.  

Previous work by Lortz\cite{Lortz} and Connor et al\cite{Connor} was 
considered for an outer flux surface that tends to a
separatrix with an X-point.  
Like Laval et al\cite{Laval} their work predicts the Peeling mode to be
unstable if there is a positive current at the plasma's edge, and it  
also finds a well behaved radial structure for the mode.  
Interestingly, for a conventional X-point there is predicted to be a
range of small but negative edge-current for which the Peeling mode is
stable, but in the limit of an exact snowflake divertor this range
shrinks to zero size - a qualitative difference between a
conventional and a snowflake divertor.   
A limitation of the Lortz calculation is that it
approximates the equilibrium quantities as constant on the length
scale of the plasma instability, this is not necessarily the case for
$q$ or $q'$ near a separatrix, and therefore the results 
when applied to a separatrix case should be
treated with caution. 
Likewise, as noted in Part (I), there are potential limitations to the
high-$n$ ordering form of $\delta W$ used here, and this should be
investigated in future work.

Thus we have developed a simple model for the Peeling mode, and found 
that despite $\delta W <0$, the growth rate $\gamma$ tends to
zero as the outermost flux surface tends to a separatrix with an
X-point.  
As the outermost flux surface approaches a separatrix, the growth rate
falls with $\ln(\gamma/\gamma_A) = - \frac{1}{2} \ln (q'/q)$; this has
subsequently been confirmed with ELITE (S. Saarelma, private
communication), leading us to believe that the effect of a separatrix
on the high toroidal mode number ideal MHD model is now understood. 
The ideal MHD prediction of marginal
stability at the separatrix means that other non-ideal terms
such as resistivity, non-linear terms, or terms neglected in the
high-$n$ analysis, will play a role in determining the
eventual stability. 
In general it  is hoped that 
the methods and results contained in this paper will provide new
tools for studying plasmas in separatrix geometries, and have
potential applications in future studies of plasma stability and more
generally outside of plasma physics.

\begin{acknowledgments}
Thanks to Jack Connor for suggesting that a conformal
transformation might help, for suggesting the method of a saddle-point
approximation, and  for many helpful discussions and suggestions.  
Thanks to Jim Hastie and Chris Gimblett, for helpful discussions and
encouragement.  
Thanks to A. Thyagaraja for pointing me to
Milne-Thompson's book (where I encountered the Karman-Trefftz
transformation).   
Thanks to Samuli Saarelma for calculations with ELITE, and Tim Hender
for reading and commenting on this paper.  
Thanks to D.D. Ryutov for questioning how stability might be different
with a snowflake divertor, and directing me to Ref. \cite{Ryutov}. 
This work was jointly funded by the United Kingdom Engineering and
Physical Sciences Research Council, and by the European Community under
the contract of Association between EURATOM and UKAEA. The views and
opinions expressed herein do not necessarily reflect those of the
European Commission.
\end{acknowledgments}


\section{$w'(z)$ near the X-point}\label{w'xpt}

To obtain $w'(z)$ we differentiate both sides of Eq. 
\ref{KTtransform} with
respect to $z$, and rearrange the resulting expression to get  
\begin{equation}
w'(z) = \left( w(z) + nl \right)^2 
\frac{ \left( z-l \right)^{n-1} }{ \left( z+ l \right)^{n+1} } 
\end{equation}
and hence 
\begin{equation}\label{w'a1}
\left| w'(z) \right|^2 = \left| w(z) +nl \right|^4 
\frac{ \left| z- l \right|^{2(n-1)} }{\left| z+ l \right|^{2(n+1)}} 
\end{equation}

We have deliberately obtained an implicit expression for $
\left|w'(z)\right|^2$, with $w'(z)$ given in terms of $w(z)$. 
This is because one cannot simply expand $w'(z)$ in powers of
$\epsilon$, because the expansion will give the incorrect answer as
$\epsilon \rightarrow 0$ compared with the exact result for $\epsilon$
small but non-zero (i.e. $\epsilon \neq 0$ is a ``singular
perturbation'').  
Instead by obtaining $|w'(z)|^2$ implicitly in the form given by
Eq. \ref{w'a1}, we need solely be careful with the term
$|z-l|^{2(n-1)}$, because $|w+nl|^4$ and $|z+l|^{2(n+1)}$ are well
behaved when expanded in $\epsilon$, as $\epsilon \rightarrow 0$.  
Near the X-point, 
\begin{equation}
\begin{array}{l}
\left| w(z) + nl \right|^4 = (2nl)^4 + \mbox{O} \left( \epsilon \right) 
\\
\left| z + l \right|^{2(n+1)} = (2l)^{2(n+1)} + \mbox{O} \left(
\epsilon \right)  
\end{array}
\end{equation}
We need to be more careful with $|z-l|^{2(n-1)}$, that with $z=-a +
(a+l-\epsilon)e^{i\alpha}$ gives  
\begin{equation}
\begin{array}{ll}
\left| z - l \right|^2 &= 
\left( (a+l-\epsilon)\cos(\alpha) -a +  l \right)^2 + (a+l+\epsilon)^2
\sin^2(\alpha)  
\\
&= 2(a+l)(a+l-\epsilon) \left[ 1 - \cos(\alpha) +
  \frac{\epsilon^2}{2(a+l)(a+l-\epsilon)} \right]  
\\
& \simeq 2a^2 \left[ 1 - \cos(\alpha) + \frac{\epsilon^2}{2a^2} \right] 
\end{array} 
\end{equation}
where we have retained the term that provides the singular
perturbation that prevents $(z-l)^2$ becoming zero for $\epsilon \neq
0$, but neglected all the lower order terms that modify the answer by
of order $\epsilon/a$ and $l/a$. In principle some plasma
cross-sections might require the retention of terms of order $l/a$,
but here we neglect them so as to keep algebraic details to a minimum.  

Therefore at leading order we have
\begin{equation}
\left| w'(z) \right|^2 
\simeq 
\frac{(2nl)^4}{(2l)^{2(n+1)}} 
(2a^2)^{(n-1)} 
\left( 1 - \cos ( \alpha ) + \frac{\epsilon^2}{2a^2} \right)^{(n-1)} 
\end{equation}
that for the case we are most interested in here with $n=3/2$
(corresponding to an X-point with a $\pi/2$ interior angle), we have  
\begin{equation}
\left| w'(z) \right|^2 
\simeq 
\frac{(3l)^4}{(2l)^5} 
\left( 2a^2 \right)^{1/2} 
\left( 1-\cos (\alpha ) + \frac{\epsilon^2}{2a^2} \right)^{1/2} 
= \frac{1}{\sqrt{2}} 
\left( \frac{3}{2} \right)^4 
\left( \frac{a}{l} \right) 
\sqrt{ 1 -\cos (\alpha ) + \frac{\epsilon^2}{2a^2} } 
\end{equation}
which is Eq. \ref{w'zsq}.

\section{Some identities involving complex numbers}\label{complexidentities}

In the following we write $a=a_x+ia_y$, $b=b_x+ib_y$, $c=c_x+ic_y$,
and $d=d_x+id_y$, and remind the reader that the dot product refers to
the sum of, the product of the real parts plus the product of the
imaginary parts. For example $a.b=a_xb_x+a_yb_y$. Multiplication is as
usual for complex numbers, for example $ab=a_xb_x-a_yb_y + i (a_xb_y +
a_y b_x)$.  Then we find 
\begin{equation}
\begin{array}{ll}
ab . cd &= \left[ (a_xb_x -a_yb_y)+i(a_xb_y+a_yb_x) \right] . 
\left[ (c_xd_x - c_yd_y)+i(c_xd_y +c_yd_x) \right]
\\
&= (a_xb_x -a_yb_y)(c_xd_x - c_yd_y) + 
   (a_xb_y +a_yb_x)(c_xd_y + c_yd_x)
\\
&= a_x \left[ b_x(c_xd_x-c_yd_y) + b_y(c_xd_y+c_yd_x) \right] + 
a_y \left[ b_x(c_xd_y+c_yd_x) -b_y(c_xd_x-c_yd_y) \right] 
\\
&= a_x \left[ d_x(b_xc_x+b_yc_y) + d_y(b_yc_x-c_yb_x) \right] + 
a_y \left[ d_y(b_xc_x+b_yc_y) + d_x(b_xc_y-b_yc_x) \right]
\\
&= (a_xd_x+a_yd_y)(b_xc_x+b_yc_y) + 
(a_xd_y-a_yd_x)(b_yc_x-c_yb_x)
\\
&= (a.d)(b.c) + (ia.d)(ic.b)  
\end{array}
\end{equation}
and similarly for $a.bc$,
\begin{equation}
\begin{array}{ll}
a.bc &=
 a.\left( ( b_xc_x-b_yc_y ) + i (b_xc_y+b_yc_x) \right) 
\\&=
 a_xb_xc_x - a_xb_yc_y + a_yb_xc_y + a_yb_yc_x 
\\&=
 c_x (a_xb_x +a_yb_y) + c_y (a_yb_x - a_xb_y ) 
\\&=
 (c_x + i c_y) . \left( (a_xb_x+a_yb_y) + i (a_yb_x-a_xb_y) \right)
\\&=
 (c_x + i c_y) . \left( a_x (b_x-ib_y) + a_y (b_y+ib_x) \right) 
\\&=
 (c_x + i c_y) . \left( a_x (b_x-ib_y) - ia_y (-b_x+ib_y) \right) 
\\&=
 c . \left( (a_x+ia_y) (b_x-ib_y) \right) 
\\&=
 c . a\bar{b} = a\bar{b} . c
\end{array}
\end{equation}

\section{2nd term is order $\frac{1}{m}$ smaller than $\delta
  W_V$}\label{app1}  
 
Here it is shown that the second term in Eq. \ref{dWV4}, here written
as $\delta W_G$, is of order $1/m$ smaller than $\delta W_V$.  
The term we are interested in is
\begin{equation}\label{dWG}
\delta W_G = \pi \Delta \left( \frac{m}{nq} \right) \xi_m^* \oint
d\alpha \frac{r_aI}{B^2} e^{-im\theta(\alpha) +in\phi}  
\frac{i}{n} \left[ 
\left( n_z . B_z \right) n_z . \nabla_z \left(
\frac{\left|B_{pz}\right|}{\left|w'(z)\right|^2} \right)  
\right]
\end{equation}
Using Eq. \ref{w'zsq} and Eq. \ref{Bpzalpha} we get 
\begin{equation}\label{A4}
\begin{array}{ll}
n_z.\nabla_z \left( \frac{\left| B_{pz} \right| }{\left| w'(z)
  \right|^2} \right)  
&= -B_{p0} \frac{\partial}{\partial \epsilon} \left[  2\sqrt{2} \left(
  \frac{2}{3} \right)^4 \frac{l}{a} \sqrt{ 1 - \cos(\alpha) +
  \epsilon^2/2a^2} \right]  
\\
&= -B_{p0} \left( \frac{\epsilon}{a^2}  \right)\sqrt{2} \left(
  \frac{2}{3} \right)^4  
\left( \frac{l}{a} \right) 
 \frac{1}{\sqrt{ 1 -\cos(\alpha) + \epsilon^2/2a^2}} 
\\
&= - \left( \frac{\epsilon}{a^2} \right) c_a \frac{R^2B_{p0}^2}{Ir_a}
  q\frac{\partial \theta}{\partial \alpha}  
\end{array}
\end{equation}
where $c_a=\left[2\sqrt{2}\left(\frac{2}{3}\right)^4 \left(
  \frac{l}{a} \right)\right]^2$ and is  a constant.  
Substituting the above Eq. \ref {A4} into Eq. \ref{dWG}, gives 
\begin{equation}
\delta W_G = \pi \Delta \left( \frac{m}{nq} \right) \xi_m^*
\left( \frac{-\epsilon}{a^2} \right) c_a \frac{iq}{n}
\oint d\alpha 
\frac{R^2B_{p0}^2}{B^2} e^{-im\theta(\alpha) +in\phi} 
 \left[ 
n_z . B_z \frac{\partial \theta}{\partial \alpha}  
\right]
\end{equation}
Integrating by parts then gives 
\begin{equation}\label{e5}
\delta W_G = \pi \Delta \left( \frac{m}{nq} \right) \xi_m^*
\left( \frac{-\epsilon}{a^2} \right) c_a \frac{q}{nm}
\oint d\alpha 
\frac{R^2B_{p0}^2}{B^2} e^{-im\theta(\alpha) +in\phi} 
 \left[ 
\frac{\partial n_z . B_z }{\partial \alpha}  
\right]
\end{equation}
Using $n_z.B_z=\frac{\partial V}{\partial r}$ and Eq. \ref{Vvac}
we get 
\begin{equation}
\left. \frac{\partial}{\partial \alpha} \left( n_z . B_z \right)
\right|_{r_a} =  
\sum_p \frac{-ip|p|a_p}{r_a} e^{ip\alpha -in\phi} 
\end{equation}
Substituting this into \ref{e5} we get 
\begin{equation}\label{e6}
\delta W_G = \pi \Delta \left( \frac{m}{nq} \right) \xi_m^*
\left( \frac{-\epsilon}{a^2} \right) c_a \frac{q}{nm} \frac{R^2B_{p0}^2}{B^2}
\sum_{p\neq0} \frac{-a_p|p|ip }{r_a} 
\oint d\alpha \mbox{ }
 e^{-im\theta(\alpha) +in\phi} 
\end{equation}
Where the poloidal dependence in $\frac{R^2}{B^2}$ has been neglected
due to the large aspect ratio Tokamak ordering.   
Using Eq. \ref{ap} we get 
\begin{equation}
\oint e^{-im\theta +ip\alpha} d\alpha = -i a_p^* \frac{|p|}{p}
\frac{R}{\Delta \xi_m^*} 2\pi  
\end{equation}
which may be inserted into \ref{e6}, and after some cancellations gives 
\begin{equation}\label{e7}
\delta W_G = 2\pi^2R 
\left( \frac{\epsilon}{a^2} \right) c_a  \frac{R^2B_{p0}^2}{B^2} \frac{1}{n^2} 
\frac{1}{r_a} \sum_{p\neq0} |p|^2 \left| a_p \right|^2 
\end{equation}
The sum $\sum_{p\neq0} |p|^2 \left| a_p \right|^2$ may be evaluated
arbitrarily accurately, as is indicated next. We use the alternative
expression for $a_p$ given by Eq. \ref{ap1}, of  
\begin{equation}
a_p = - \frac{\Delta}{|p|} \frac{\xi_m}{R} \frac{1}{2\pi} \oint im
\theta'(\alpha) e^{im\theta -ip\alpha} d\alpha 
\end{equation}
which gives 
\begin{equation}
\begin{array}{ll}
\sum_p |p|^2 \left|a_p \right|^2 
& = \Delta^2 \frac{\left| \xi_m \right|^2}{R^2} \sum_{p\neq0} 
\left( \frac{1}{2\pi} \oint d\beta \mbox{ } m \theta'(\beta)
e^{-im\theta(\beta)+ip\beta}  \right) 
\times
\\
& \left( \frac{1}{2\pi} \oint d\alpha \mbox{ } m \theta'(\alpha)
e^{im\theta(\alpha)-ip\alpha} \right)  
\\
& = \Delta^2 \frac{\left| \xi_m \right|^2}{R^2} 
\frac{m^2}{2\pi} 
\oint d\beta \mbox{ } \theta'(\beta)  e^{-im\theta(\beta)} 
\sum_{p\neq 0} e^{+ip\beta} 
\frac{1}{2\pi} \oint d\alpha \mbox{ } \theta'(\alpha)
e^{im\theta(\alpha)}e^{-ip\alpha}  
\\
& = \Delta^2 \frac{\left| \xi_m \right|^2}{R^2} 
\frac{m^2}{2\pi} 
\oint \theta'(\beta)  e^{-im\theta(\beta)} \theta'(\beta)
e^{im\theta(\beta)} d\beta  
\\
& = \Delta^2 \frac{\left| \xi_m \right|^2}{R^2} 
\frac{m^2}{2\pi} 
\oint \left( \theta'(\beta) \right)^2 d\beta
\end{array}
\end{equation}
where in the penultimate line, we resum the Fourier series by noting
that $\theta'(\beta) e^{im\theta(\beta)}=
\sum_{p\neq 0} e^{ip\beta} \frac{1}{2\pi} \oint
e^{-ip\alpha} \theta'(\alpha) e^{im\theta(\alpha)}$. 
We 
approximate $\oint \left( \theta'(\beta) \right)^2 d\beta$, using the 
analytic expression for $\theta(\beta)$ obtained from
Eqs. \ref{theta1},  \ref{Bpzalpha}, and  \ref{w'zsq}, obtaining  
\begin{equation}
\oint \left( \theta'(\beta) \right)^2 d\beta 
= \frac{1}{q^2} \frac{1}{c_a} \frac{I^2r_a^2}{R^4B_{p0}^2}  \oint
\frac{d\beta}{1-\cos(\beta) + \epsilon^2/2a^2}  
 \simeq  \frac{1}{q^2} \frac{1}{c_a} \frac{I^2r_a^2}{R^4B_{p0}^2} 
\frac{2\pi a}{\epsilon } 
\end{equation}
Hence taking  $r_a=a+l \simeq a$ we get 
\begin{equation}
\begin{array}{ll}
\delta W_G &\simeq \left( 2\pi^2 R \frac{\epsilon}{a^2} c_a
\frac{R^2B_{p0}^2}{B^2}  
\frac{1}{n^2} \right) 
\left( \Delta^2 
\frac{\left| \xi_m \right|^2}{R^2} \frac{m^2}{2\pi} \right)
\left( 
\frac{I^2r_a^2}{R^4B_{p0}^2} \frac{1}{c_a} \frac{1}{q^2} \frac{2\pi
  a}{\epsilon}  
\right) 
\\
&= 2\pi^2 \frac{\left| \xi_m \right|^2}{R} \Delta^2
= \left( \frac{1}{m} \right) \delta W_V
\end{array}
\end{equation}
where the last line uses the expression for $\delta W_V \sim \Delta^2
m$ obtained from the main text. Hence, we may neglect this term
compared with $\delta W_V$.

\section{Modifications to Peeling Mode Stability for a ``Snowflake''
  divertor} 

The integrals that determine the stability of Peeling modes, are
dominated by the divergence in $\nu/B_p^2$ that occurs near the
X-point in the separatrix. 
This allows the integrals to be estimated by expanding the poloidal
flux functions in the vicinity of the X-point, as has been done in
Ref. \cite{Ryutov} for both a standard X-point and a ``snowflake''
divertor's X-point. 
Near a standard X-point, Eq. 15 of Ref. \cite{Ryutov} gives 
\begin{equation}\label{phiX}
\Phi = \left( \frac{\tilde{B}}{L} \right) \left( \frac{x^2-z^2}{2}
\right) 
\end{equation}
with $\tilde{B}$ having the poloidal field's dimensions and $L$ a
length scale, 
and the poloidal magnetic field given by\cite{Ryutov}
$B_x=-\frac{\partial \Phi}{\partial z}$ and $B_z=\frac{\partial
  \Phi}{\partial x}$, leading to
\begin{equation}
B_p^2 = \frac{\tilde{B}^2}{L^2} \left( x^2 + z^2 \right) 
\end{equation}
Note that near the X-point, $\nabla \psi \rightarrow R_X \nabla
\Phi$, with $R_X$ the major radius at the X-point. 
Therefore near the X-point $\frac{\partial B_p^2}{\partial \psi} =
\frac{1}{R_X} \frac{\partial B_p^2}{\partial \Phi}$, and we 
calculate $\partial B_p^2/\partial \psi$ from 
\begin{equation}
\begin{array}{ll}
\frac{\partial B_p^2}{\partial \Phi} &= 
\frac{\nabla \Phi . \nabla \left( B_p^2 \right)}{\left| \nabla \Phi
  \right|^2}  
\\
&= \left( \frac{\tilde{B}}{L} \right) 2 \frac{(x^2-z^2)}{(x^2+z^2)} 
\end{array}
\end{equation}
and use Eq. \ref{phiX} to substitute for $x^2=2 \left( \frac{\Phi
  L}{\tilde{B}} \right) + 2 z^2$, giving $\partial B_p^2/\partial
  \Phi$ at constant $\Phi$ as 
\begin{equation}
\frac{\partial B_p^2}{\partial \Phi} = 2 \left( 
\frac{\tilde{B}}{L} \right) 
\frac{\left( \Phi L/\tilde{B} \right) }{
\left( z^2 + \left( \Phi L/\tilde{B} \right) \right)}
\end{equation}
and similarly 
$B_p^2 = \left( \tilde{B}^2/L^2 \right) 2 
\left( z^2 + \Phi L/\tilde{B} \right)$. 
The minimum value of $z$ is at $x=0$, giving 
$z_{min}^2=-2\Phi L/\tilde{B}$, and 
\begin{equation}
\left. 
\frac{\partial B_p^2}{\partial \Phi} \right|_{z_{min}} 
= -2 \left( \frac{\tilde{B}}{L} \right) 
\end{equation}

For a ``snowflake'' divertor, Eq. (2) of Ref. \cite{Ryutov} gives 
\begin{equation}
\Phi = \frac{AI}{c} \left( x^2 z - \frac{z^3}{3} \right) 
\end{equation}
with I the plasma current, A has dimensions of the inverse cube of the
length scale over which the poloidal magnetic field varies near an
X-point, and c is the speed of light. 
In a similar way to the calculation for the standard X-point, this
leads to 
\begin{equation}
\frac{\partial B_p^2}{\partial \Phi}
= \left( \frac{AI}{ c} \right) 
\frac{ 4z \left( \frac{c\Phi}{AI} \right)}{
\left( \frac{c\Phi}{AI} \right) + z^3} 
\end{equation}
\begin{equation}
B_p^2 = \left( \frac{A^2I^2}{c^2} \right) 
\left( \frac{4}{3} z^2 + \left( \frac{c\Phi}{AI} \right) \frac{1}{z}
\right)^2 
\end{equation}
with $z_{min}^3 = - 3 \Phi c/AI $, giving 
\begin{equation}
\left. 
\frac{\partial B_p^2}{\partial \Phi} 
\right|_{z_{min}} 
= 
\left( \frac{AI}{c} \right) \frac{4}{3} 
\left( \frac{c\Phi}{AI} \right)^{1/3} 
\end{equation}
that tends to zero as we approach a separatrix. 
Therefore, whereas a conventional X-point leads to a Peeling mode
stability boundary for which the Peeling mode can be stable for a
range of small but non-zero negative current at the plasma's edge,
this window of stability tends to zero size for a snowflake
divertor. 

Finally, for a ``snowflake plus'' divertor\cite{Ryutov}, a similar
calculation using the Eqs. of Ref. \cite{Ryutov} finds $\partial
B_p^2/\partial \Phi \sim \sqrt{\frac{I-I_{d0}}{I_{d0}}}$ with $I$ the
current in the divertor coils and $I_{d0}$ the divertor coil current
required for an exact snowflake divertor. 
In that case the range of values of negative edge-current for which
Peeling modes are stable, tends to zero as $I \rightarrow I_{d0}$.

\end{document}